\documentclass{article}
\usepackage{jheppub} 
\usepackage[utf8x]{inputenc}	
\usepackage[english]{babel}

\usepackage[ddmmyyyy,hhmmss]{datetime}

\usepackage{multirow}

\catcode`@=11
\allowdisplaybreaks
\usepackage{amsxtra}
\usepackage{mathtools}
\usepackage{float}
\usepackage{xcolor}
\usepackage{tikz}
\usepackage{stackengine}
\usepackage{placeins}
\usepackage{booktabs}
\usepackage{macros}
\usepackage{hyperref}
\usepackage{subcaption}
\usepackage{amsmath}
\usepackage{graphicx}
\usepackage{rotating}
\usepackage{epigraph}

\usetikzlibrary{hobby,intersections,positioning} 
\usetikzlibrary{shapes.geometric}

\newcommand{\Figref}[1]{Figure~\ref{#1}}
\newcommand{\Quiver}[1]{$\mathcal Q_{\ref{#1}}$}

\newcommand{\hs}{\mathrm{HS}}

\newcommand{\pe}{\mathrm{PE}}
\newcommand{\hwg}{\mathrm{HWG}}
\usepackage{placeins}
\usetikzlibrary{calc}
\usetikzlibrary{arrows.meta, bending, patterns}

\newcommand{\hsC}[1]{\hs\left[\mathcal C\left(\text{\Quiver{#1}}\right)\right]}
\newcommand{\hsH}[1]{\hs\left[\mathcal H\left(\text{\Quiver{#1}}\right)\right]}
\newcommand{\Coul}[1]{\mathcal C\left(\text{\Quiver{#1}}\right)}
\newcommand{\Higgs}[1]{\mathcal H\left(\text{\Quiver{#1}}\right)}

\tikzset{flavour/.style={draw=none,minimum size=0.3mm,fill=white, regular polygon,regular polygon sides=4,draw}}
\tikzset{flavourr/.style={draw=none,minimum size=0.3mm,fill=red, regular polygon,regular polygon sides=4,draw}}
\tikzset{flavourb/.style={draw=none,minimum size=0.3mm,fill=blue, regular polygon,regular polygon sides=4,draw}}
\tikzset{gaugeBig/.style={inner sep=1mm,draw=none,fill=white,minimum size=2mm,circle, draw}}
\tikzset{bd/.style={circle, draw=black, inner sep=0pt, fill=black, minimum size=2mm}}
\tikzset{wd/.style={circle, draw=black, inner sep=0pt, fill=white, minimum size=2mm}}
\tikzset{Dynkin/.style={circle, draw=black, inner sep=0pt, fill=white, minimum size=2mm}}
\tikzstyle{ligne}=[draw, very thick] 
\tikzstyle{gridline}=[draw, gray] 
\tikzset{gauge/.style={circle, draw,inner sep=2.5pt}}
\tikzset{gaugeo/.style={circle, draw,inner sep=2.5pt,fill=orange}}
\tikzset{gauger/.style={circle, draw,inner sep=2.5pt,fill=red}}
\tikzset{gaugeb/.style={circle, draw,inner sep=2.5pt,fill=blue}}
\tikzset{gaugeg/.style={circle, draw,inner sep=2.5pt,fill=green}}
\tikzset{gaugegoodgreen/.style={circle, draw,inner sep=2.5pt,fill=goodgreen}}
\tikzset{gaugem/.style={circle, draw,inner sep=2.5pt,fill=magenta}}
\tikzset{hasse/.style={circle, fill,inner sep=2pt}}
\tikzset{d2/.style={circle, fill,inner sep=1.3pt}}
\tikzset{shrinky/.style={circle, fill,inner sep=1pt}}
\tikzset{sized/.style={circle, draw, inner sep=1.5pt}}
\tikzset{seven/.style={circle, draw,inner sep=3pt}}

\tikzset{gaugebl/.style={circle,draw=black,fill=black,inner sep=1.5pt}}
\tikzset{gaugeblnormal/.style={circle,draw=black,fill=black,inner sep=2.5pt}}
\tikzset{hasse/.style={circle, fill,inner sep=2pt}}
\tikzstyle{dashed_brane}=[thick, dashed]
\tikzstyle{dotted_brane}=[thick, dotted]
\tikzstyle{O3plus}=[thick, color=green]
\tikzstyle{O3minustilde}=[thick, color=blue]
\tikzstyle{O3plustilde}=[thick, color=red]
\tikzset{D5/.style={cross out, draw=black, minimum size=7, inner sep=0pt, outer sep=0pt}, cross/.default={1pt}}
\tikzset{flavor/.style={regular polygon,regular polygon sides=4,inner sep=2.5pt, label = {}, draw}}
\tikzset{redflavor/.style={regular polygon,regular polygon sides=4,inner sep=2.5pt, color=red, label = {}, draw}}
\tikzset{redgauge/.style={inner sep=1mm,color=red,draw=none,minimum size=2mm,circle, draw}}
\tikzset{blueflavor/.style={regular polygon,regular polygon sides=4,inner sep=2.5pt, color=blue, label = {}, draw}}
\tikzset{bluegauge/.style={inner sep=1mm,color=blue,draw=none,minimum size=2mm,circle, draw}}
\allowdisplaybreaks

\usepackage{todonotes}

\title{Orthosyplectic Quotient Quiver Subtraction II: Framed Quivers}

\author{Sam Bennett,}
\author{Amihay Hanany,}
\author{and Guhesh Kumaran}

\affiliation{Abdus Salam Centre for Theoretical Physics, Imperial College London,\\ Prince Consort Road
London, SW7 2AZ, UK}
\emailAdd{samuel.bennett18@imperial.ac.uk}
\emailAdd{a.hanany@imperial.ac.uk}
\emailAdd{guhesh.kumaran18@imperial.ac.uk}

\preprint{Imperial/TP/25/AH/03}
\abstract{The technique of \textit{orthosymplectic quotient quiver subtraction} is introduced for framed orthosymplectic quivers. This involves subtracting an \textit{orthosymplectic quotient quiver} from a framed orthosymplectic $3d\;\mathcal N=4$ quiver gauge theory which has the effect of gauging an $\sorm(n)$ or $\sprm(n)$ subgroup of the IR Coulomb branch global symmetry with complete Higgsing. The orthosymplectic quotient quivers take the form of magnetic quivers for class $\mathcal S$ theories on cylinders with (twisted) maximal punctures of simply laced algebras, similar to the case of unitary quotient quiver subtraction. This gives a set of quotient quivers for all classical groups. Notably, quotient quiver subtraction for framed and unframed orthosymplectic quivers are different procedures.

}
\begin{document}
\maketitle
\flushbottom

\section{Introduction}
A longstanding problem in the study of $3d\;\mathcal N=4$ gauge theories is finding a systematic way to gauge subgroups of IR Coulomb branch global symmetries (henceforth dropping ``IR"). The challenge is mainly due to non-perturbative effects \cite{Intriligator:1996ex,Seiberg:1996bs} and (upon quantisation) the appearance of dressed monopole operators, which render traditional Lagrangian techniques ineffective.

Significant progress in this direction has occurred in the past couple of years by focusing on $3d\;\mathcal N=4$ \textit{quiver} gauge theories with both unitary \cite{Hanany:2023tvn,Hanany:2024fqf} and orthogonal and symplectic gauge groups \cite{Bennett:2024llh}. These methods turn the operation of gauging subgroups of the Coulomb branch global symmetries into simple quiver combinatorics, bypassing the complications of non-perturbative physics.

The study of $3d\;\mathcal N=4$ quivers, and in particular their Coulomb branches, reaches beyond three dimensions to theories with eight supercharges in $d=4,5,6$. The notion of a \textit{magnetic quiver} \cite{Cabrera:2019izd,Cabrera:2019dob} gives a description of the Higgs branches of theories with eight supercharges as a moduli space of dressed monopole operators. Many phenomena to do with the appearance of additional massless states, at certain values of the gauge coupling, are encapsulated in the magnetic quiver where other techniques are insufficient for the analysis. As a result, developing techniques on $3d\;\mathcal N=4$ Coulomb branches gives insight into non-trivial physics in other dimensions.

The problem of gauging Coulomb branch isometry subgroups also has intimate connections with the geometry of \textit{symplectic singularities} \cite{2000InMat.139..541B}. Both the Higgs branch and the Coulomb branch of $3d\;\mathcal N=4$ theories are hyper-Kähler cones \cite{Hitchin:1986ea}. Gauging a subgroup of the isometry of a moduli space of vacua (of either the Higgs branch or Coulomb branch) introduces additional F-term relations that project out some of the previously gauge-invariant operators. Geometrically, this is the action of a hyper-K\"ahler quotient, where the F-term plays the role of a complex moment map. In cases where there is complete Higgsing, this can be seen using the Hilbert series of the Higgs \cite{Benvenuti:2006qr} and Coulomb branches \cite{Cremonesi:2013lqa}.
Since 3d $\mathcal{N}=4$ theories have both Higgs and Coulomb branches, gauging a subgroup of the isometry of one branch has the effect of ungauging this subgroup of the isometry of the other branch. 

The diagrammatic techniques developed in \cite{Hanany:2023tvn,Hanany:2024fqf,Bennett:2024llh} were inspired by the physics of theories with eight supercharges in four and six dimensions. Using magnetic quivers \cite{Cabrera:2019izd,Cabrera:2019dob}, the action of gauging a Higgs branch isometry in an electric theory can be restated as the subtraction of a \textit{quotient quiver} from its magnetic counterpart or a \textit{polymerisation} of magnetic quivers.\footnote{The technique of \textit{quiver polymerisation} \cite{Hanany:2024fqf} follows from gluing maximal punctures of $4d\;\mathcal N=2$ class $\mathcal S$ theories \cite{Gaiotto:2009we,Benini:2010uu}.} 

For an electric theory with unitary gauge nodes, customarily termed a `unitary quiver', the quotient quiver is the magnetic quiver for the class $\mathcal S$ theory on a cylinder with two maximal $A$-type punctures.

For unitary (magnetic) quivers, there is an overall centre-of-mass $\urm(1)$ which may be fixed. This comes from a uniform shift of all magnetic charges of the centre of the gauge groups. Fixing this $\urm(1)$ turns an unframed quiver into a framed quiver  -- diagramatically, this is simply the addition of a flavour node to a quiver that did not previously contain one. The unframed quiver describes an equivalence class of framed quivers which have the same Coulomb branch (up to discrete actions \cite{Hanany:2020jzl}). In the mathematics literature, the (un)fixing of the centre-of-mass $\urm(1)$ is referred to as the Crawley-Boevey trick \cite{Crawley-Boevey2001GeometryQuivers}. In the unitary case, it turns out that the quotient quiver subtraction procedures on framed and unframed theories are virtually the same.

Unfortunately, these shared combinatorics do not extend to \textit{orthosymplectic} quivers, formed out of orthogonal and symplectic gauge groups. The magnetic lattice for these theories has a fixed origin which prevents an overall $\urm(1)$ shift of the charges. As a result, unframed orthosymplectic quivers cannot be thought of as an equivalence class of framed orthosymplectic quivers. 

So far, quotient quivers have been found for the groups $\surm(2)$, $\surm(3)$, $G_2$, and $\sorm(7)$ for unframed orthosymplectic quotient quiver subtraction. The derivation in \cite{Bennett:2024llh} involved gauging flavour subgroups of $6d\;\mathcal N=(1,0)\;\sprm(k)$ gauge theory at infinite coupling with $2k+8$ flavours and comparing the magnetic quivers, derived from the Type IIA brane systems, before and after. The resulting quotient quivers were confined to the four groups above due to the restrictions anomaly cancellation places on matter content. The quivers for six dimensional theories including the gauge group $\sorm(n)$ for $n\geq8$ require multiple types of matter to be coupled and so the quiver has bifurcations and it is unclear how to realise them in brane systems. Hence, the quotient quivers that were found in \cite{Bennett:2024llh} populated a puzzling sequence that resisted attempts at generalisation.

This work presents quotient quivers for framed orthosymplectic theories. These take the form of magnetic quivers for class $\mathcal S$ theories on a cylinder with maximal (twisted) $A$-type or $D$-type punctures. The quotient quivers lie in a one-parameter family with the effect of gauging $G=\sorm(n)$ or $\sprm(n)$ subgroups of the Coulomb branch global symmetry.


In \cite{Sperling:2021fcf} a collection of brane systems and magnetic quivers were proposed for moduli spaces which are a product of two symplectic singularities. In particular, this construction uses $\text{ON}^{-}$ planes \cite{Hanany:2000fq}, which have the effect of doubling the global symmetry. Examples include brane systems and magnetic quivers for the Higgs branch of a product of framed one node quivers. In the electric theory, gauging a diagonal flavour symmetry subgroup results in a three-node framed quiver. Three-node framed quivers may be interpreted using two inequivalent brane systems under the isomorphism $A_3\simeq D_3$. One interpretation is that the gauge nodes form an $A_3$ Dynkin diagram and so the brane system consists of a linear set of segments where light D-branes can end. Another valid interpretation is that the gauge nodes form a $D_3$ Dynkin diagram and so the brane system includes an $\text{ON}^{-}$ plane to yield the bifurcation. Both brane systems give different magnetic quivers and it is expected that they flow from different UV origins to the same IR fixed point. This can be checked by computing the Higgs branch and Coulomb branch Hilbert series for each quiver and showing that they match.

The structure of this paper is as follows. Section \ref{sec:taxonomy} offers a summary of the various processes that trade under the name `quiver subtraction', making explicit the essential differences between them. Section \ref{sec:derivs} introduces the set of quotient quivers considered in this work and compares them to the unitary quotient quivers given in \cite{Hanany:2023tvn}. Section \ref{sec:rules} lists the rules for orthosymplectic quotient quiver subtraction, which differ from those for both the unitary and unframed orthosymplectic cases (other rules, such as the junction rule, make an appearance). Section \ref{sec:simple_examples} presents some simple examples of the algorithm in action, which are extended in Section \ref{sec:examples} -- the junction rule is tested explicitly in Section \ref{sec:testing_junction_rule}. An outlook of the work done here is presented in Section \ref{sec:outlook}. Appendix \ref{sec:so(4n+2)_(4n+1)_flavours} details an attempt of $\sprm(n)$ quotient quiver subtraction.

\subsection{Quiver, Hilbert series, and brane conventions}
All theories in this paper are in three dimensions with eight supercharges. Red nodes denote gauge nodes of $\sorm$ groups (unless explicitly specified as $\orm$), while blue nodes denote $\sprm$ groups with the notation $\sprm(1)\simeq\surm(2)$. Gauge nodes will be labelled by their rank.

The term `framed quiver' refers to the presence of a (square) flavour node. In this sense, a `framed quotient quiver' refers to a quotient quiver that is subtracted from a framed target quiver. Unframed quotient quivers are similarly subtracted from unframed target quivers. Note that the quotient quivers themselves are always unframed.

Quivers will be denoted by $\mathcal Q$ labelled by a figure number or equation number in the subscript.

Any particularly lengthy palindromials in the numerator of the Hilbert series will be abbreviated with an ellipsis after the first half. Any Plethystic Logarithms \cite{Feng:2007ur} of a Hilbert series will be presented perturbatively.

\begin{table}[h]
    \centering
\begin{tabular}{ccccccccccc}
  \toprule
  \\[-1em]
   & $x^{0}$ & $x^{1}$ & $x^{2}$ & $x^{3}$ & $x^{4}$ & $x^{5}$ & $x^{6}$ & $x^{7}$ & $x^{8}$ & $x^{9}$\\
  \midrule
  NS5/ON & \checkmark & \checkmark & \checkmark & \checkmark & \checkmark & \checkmark & & & &\\ 
  \midrule
  D5/O5 & \checkmark & \checkmark & \checkmark & & & & & \checkmark & \checkmark & \checkmark\\ 
  \midrule
  D3/O3 & \checkmark & \checkmark & \checkmark &  & & & \checkmark &  & & \\ 
  \bottomrule
\end{tabular}
\caption{The Type IIB configurations in this paper consist of D3-branes suspended between D5-branes and/or NS5-branes, together with O3, O5 and $\text{ON}^{-}$ orientifold planes. The spacetime dimensions occupied by each object is checkmarked.}
\label{tab:braneocc}
\end{table}
\begin{table}[H]
\ra{1.5}
    \centering
    \begin{tabular}{cccc}
    \toprule
          Orientifold Plane & Brane Diagram & Electric Gauge Algebra & Magnetic Gauge Algebra\\ \midrule
          O3$^{-}$ & & $\sorm(2k)$ & $\sorm(2k)$
          \\\midrule $\widetilde{\text{O3}^{-}}$ & \begin{tikzpicture}
              \draw[-] (-1,0) -- (1,0);
          \end{tikzpicture} & $\sorm(2k+1)$ & $\sprm(k)$\\ \midrule
          O3$^{+}$ & \begin{tikzpicture}
              \draw[dashed_brane] (-1,0) -- (1,0);
          \end{tikzpicture} & $\sprm(k)$ & $\sorm(2k+1)$ \\ \midrule
          $\widetilde{\text{O3}^{+}}$ & \begin{tikzpicture}
              \draw[dotted_brane] (-1,0) -- (1,0);
          \end{tikzpicture} & $\sprm(k)$ & $\sprm(k)$\\
          \bottomrule
    \end{tabular}
    \caption{The identification of gauge algebras from O3-planes and brane diagram conventions. Switching from the electric to magnetic gauge algebra involves S-duality.}
    \label{orientifold_table}
\end{table}
A summary of the spacetime occupations of branes are given in Table \ref{tab:braneocc} and the various O3 planes that will appear are given in Table \ref{orientifold_table}. For instance, the $x^{3,4,5}$ directions will appear into the page, while the $x^{7,8,9}$ are vertical and the $x^{6}$ direction is horizontal.
\subsection{The taxonomy of orthosymplectic quiver subtraction}
\label{sec:taxonomy}
Over the past several years, various notions of `quiver subtraction' have been employed to capture different operations on quiver gauge theories with eight supercharges. At present, the term means one of two things. Generically, `quiver subtraction' refers to the quiver combinatorics used to determine the symplectic stratification of the Higgs/Coulomb branch of a given quiver gauge theory. Although this technique finds success when applied to unitary quiver theories, the particularities of orthosymplectic gauge groups ensure that proposals for generic subtraction algorithms are unable to capture the full range of Higgsing patterns on both the Coulomb and Higgs branches \cite{Bourget:2021xex}. Recent advances in this direction include the \textit{decay} and \textit{fission} algorithms \cite{Bourget:2023dkj,Bourget:2024mgn,Lawrie:2024wan}. 

The second type of quiver combinatorics, termed `quotient quiver subtraction', gauges subgroups of the Coulomb branch global symmetry of a 3d $\mathcal{N}=4$ quiver gauge theory in terms of a subtraction operation on the quiver \cite{Hanany:2023tvn,Bennett:2024llh}. In this case, the subtraction of quivers no longer corresponds to a choice of vacuum and often results in theories which do not fall into the typical Higgsing pattern of the original theory.

\section{Quotient quivers for framed orthosymplectic theories}
\label{sec:derivs}
As introduced in \cite{Hanany:2023tvn}, gauging (with complete Higgsing) an $\surm(n)$ subgroup of the Coulomb branch global symmetry of an unframed unitary quiver can be performed by subtracting the quotient quiver given in \eqref{quiv:unitary_quotient_quiver}. This quotient quiver is interpreted as the magnetic quiver for the class $\mathcal S$ theory on a cylinder with two maximal $A_{n-1}$ punctures.
\begin{equation}
    \begin{tikzpicture}
        \node[gauge, label=below:$1$] (1l) at (0,0){};
        \node[gauge, label=below:$2$] (2l) at (1,0){};
        \node[] (cdotsl) at (2,0){$\cdots$};
        \node[gauge, label=below:$n$] (n) at (3,0){};
        \node[] (cdotsr) at (4,0){$\cdots$};
        \node[gauge, label=below:$2$] (2r) at (5,0){};
        \node[gauge, label=below:$1$] (1r) at (6,0){};

        \draw[-] (1l)--(2l)--(cdotsl)--(n)--(cdotsr)--(2r)--(1r);
    \end{tikzpicture}
    \label{quiv:unitary_quotient_quiver}
\end{equation}
In analogy with this construction, it is natural to consider magnetic quivers for class $\mathcal S$ theories on cylinders with maximal (twisted) punctures of simply laced algebras. Indeed, calculations show that these are quotient quivers for framed orthosymplectic theories, introduced below.

The quivers in Table \ref{table:framed_ortho_quotient_quivers} comprise the various quotient quivers that, upon their subtraction from a a framed orthosymplectic quiver, gauge a Coulomb branch global symmetry subgroup. The quivers presented here correspond to the $D/B/C$ families of classical Lie groups; at present, it is not known whether gauging a Coulomb branch global symmetry subgroup of exceptional type admits a \hyperref[sec:rules]{quotient quiver subtraction}. This differs from the unframed case, in which the process of gauging a $G_2$ Coulomb branch global symmetry subgroup finds a description as the quotient quiver given in Table \ref{table:unframed_quotient_quivers}.
\begin{table}[h]
        \centering
        \begin{tabular}{cc}
        \toprule
             $G$ & Framed Quotient Quiver\\
             \midrule
             $\sorm(2n)$ & \raisebox{-0.5\height}{\begin{tikzpicture} 
        \node (1) [gauger, label=below:{$D_1$}] at (0,0) {};
        \node (2) [gaugeb, label=below:{$C_1$}] at (1,0) {};
        \node (3) [gaugeb, label=below:{$C_{n-1}$}] at (4,0) {};
        \node (4) [gauger, label=below:{$D_{n}$}] at (5,0) {};
        \node (7) [gaugeb, label=below:{$C_{n-1}$}] at (6,0) {};
        \node (8) [gaugeb, label=below:{$C_1$}] at (9,0) {};
        \node (9) [gauger, label=below:{$D_1$}] at (10,0) {};
        \draw (1) -- (2);
        \draw (3) -- (4);
        \draw (2) -- (2,0);
        \draw (3) -- (3,0);
        \draw (4)--(7)--(7,0);
        \draw (8,0)--(8)--(9);
        \node at (2.54, -0.03) [scale=2]{$\cdots$} {};
        \node at (7.54, -0.03) [scale=2]{$\cdots$} {};
        \end{tikzpicture}}\\
             \midrule
             $\sorm(2n+1)$ &      \raisebox{-0.5\height}{\begin{tikzpicture} 
        \node (1) [gauger, label=below:{$D_1$}] at (0,0) {};
        \node (2) [gaugeb, label=below:{$C_1$}] at (1,0) {};
        \node (3) [gauger, label=below:{$D_{n}$}] at (4,0) {};
        \node (4) [gaugeb, label=below:{$C_{n}$}] at (5,0) {};
        \node (7) [gauger, label=below:{$D_{n}$}] at (6,0) {};
        \node (8) [gaugeb, label=below:{$C_1$}] at (9,0) {};
        \node (9) [gauger, label=below:{$D_1$}] at (10,0) {};
        \draw (1) -- (2);
        \draw (3) -- (4);
        \draw (2) -- (2,0);
        \draw (3) -- (3,0);
        \draw (4)--(7)--(7,0);
        \draw (8,0)--(8)--(9);
        \node at (2.54, -0.03) [scale=2]{$\cdots$} {};
        \node at (7.54, -0.03) [scale=2]{$\cdots$} {};
        \end{tikzpicture}}\\
             \midrule
             $\sprm(n)$ & \raisebox{-0.5\height}{\begin{tikzpicture} 
        \node (1) [gauger, label=below:{$B_0$}] at (0,0) {};
        \node (2) [gaugeb, label=below:{$C_1$}] at (1,0) {};
        \node (3) [gaugeb, label=below:{$C_{n}$}] at (4,0) {};
        \node (4) [gauger, label=below:{$B_{n}$}] at (5,0) {};
        \node (7) [gaugeb, label=below:{$C_{n}$}] at (6,0) {};
        \node (8) [gaugeb, label=below:{$C_1$}] at (9,0) {};
        \node (9) [gauger, label=below:{$B_0$}] at (10,0) {};
        \draw (1) -- (2);
        \draw (3) -- (4);
        \draw (2) -- (2,0);
        \draw (3) -- (3,0);
        \draw (4)--(7)--(7,0);
        \draw (8,0)--(8)--(9);
        \node at (2.54, -0.03) [scale=2]{$\cdots$} {};
        \node at (7.54, -0.03) [scale=2]{$\cdots$} {};
        \end{tikzpicture}}\\
        \bottomrule
    \end{tabular}
    \caption{The quotient quivers for framed orthosymplectic theories. $G$ denotes the Coulomb branch global symmetry to be gauged.}
    \label{table:framed_ortho_quotient_quivers}
\end{table}
\paragraph{\boldmath$\sorm(2n)$\unboldmath}
The $\sorm(2n)$ quotient quiver is given in the first row of Table \ref{table:framed_ortho_quotient_quivers}.
This is the magnetic quiver for the class $\mathcal S$ theory on a cylinder with two maximal $D_n$ punctures. The Higgs branch dimension of the class $\mathcal S$ theory is $n(2n-1) = \text{dim}\left(\sorm(2n)\right)$ which is the total rank of the gauge nodes in the magnetic quiver. 
\paragraph{\boldmath$\sorm(2n+1)$\unboldmath}
The $\sorm(2n+1)$ quotient quiver is given in the second row of Table \ref{table:framed_ortho_quotient_quivers}. This is the magnetic quiver for the class $\mathcal S$ theory on a cylinder with two maximal $\mathbb Z_2$ twisted $D_{n+1}$ punctures. The Higgs branch dimension of the class $\mathcal S$ theory is $n(2n+1) = \text{dim}\left(\sorm(2n+1)\right)$ which is the total rank of the gauge nodes in the magnetic quiver. Note that the central node of the magnetic quiver is $C_n$ which is the Langlands or GNO dual of $B_n$.
\paragraph{\boldmath$\sprm(n)$\unboldmath}
The (conjectural) $\sprm(n)$ quotient quiver is given in the third row of Table \ref{table:framed_ortho_quotient_quivers}. This is the magnetic quiver for the class $\mathcal S$ theory on a cylinder with two $\mathbb Z_2$ twisted maximal $A_{2n-1}$ punctures. The Higgs branch dimension of the class $\mathcal S$ theory is $n(2n+1) = \text{dim}\left(\sprm(n)\right)$ which is the total rank of the gauge nodes in the magnetic quiver. Note that the central node of the magnetic quiver is $B_n$ which is the Langlands or GNO dual of $C_n$. This quotient quiver remains conjectural owing to the fact that long legs of B-C-type have underbalanced C-type gauge nodes \cite{Gaiotto:2008ak}. Present techniques are unable to find Coulomb branch Hilbert series for such theories, which are necessary to verify the procedure. However, such a subtraction can still be seen diagrammatically as illustrated for $\orm(4n+2)$ SQCD with $4n+1$ flavours as shown in Appendix \ref{sec:so(4n+2)_(4n+1)_flavours}.

This gives a set of quotient quivers for all classical groups, all of them associated to class $\mathcal S$ theories on a cylinder. Although the Coulomb branch Hilbert series of these quotient quivers cannot be computed using the monopole formula \cite{Cremonesi:2013lqa}, it is conjectured that the moduli space is $T^*G$ where $G=\surm(n),\;\sorm(2n),\;\sorm(2n+1),\;\sprm(n)$ \cite{Benini:2010uu,Chacaltana:2010ks,Chacaltana:2011ze,Chacaltana:2012ch,Chacaltana:2013oka}.

\subsection{The case of $G=\mathrm{Sp}'(n)$}
One may expect that there is also a framed quotient quiver for $G=\mathrm{Sp}'(n)$ to complete the quartet of classical groups. The group $\mathrm{Sp}'(n)$ has made several appearances in the study of string backgrounds with orientifold planes \cite{Hanany:2001iy,Feng:2000eq}.

A natural framed quotient quiver for $\mathrm{Sp}'(n)$ may be proposed by taking the $\sprm(n)$ quotient quiver in the last row of Table \ref{table:framed_ortho_quotient_quivers} and making the central node $C_n$, shown below
\begin{equation}
    \begin{tikzpicture} 
        \node (1) [gauger, label=below:{$B_0$}] at (0,0) {};
        \node (2) [gaugeb, label=below:{$C_1$}] at (1,0) {};
        \node (3) [gauger, label=below:{$B_{n-1}$}] at (4,0) {};
        \node (4) [gaugeb, label=below:{$C_{n}$}] at (5,0) {};
        \node (7) [gauger, label=below:{$B_{n-1}$}] at (6,0) {};
        \node (8) [gaugeb, label=below:{$C_1$}] at (9,0) {};
        \node (9) [gauger, label=below:{$B_0$}] at (10,0) {};
        \draw (1) -- (2);
        \draw (3) -- (4);
        \draw (2) -- (2,0);
        \draw (3) -- (3,0);
        \draw (4)--(7)--(7,0);
        \draw (8,0)--(8)--(9);
        \node at (3, 0){$\cdots$} {};
        \node at (8,0) {$\cdots$} {};
        \end{tikzpicture}
\end{equation}

The total rank of the gauge nodes is $n(2n-1)=\text{dim}\left(\sorm(2n)\right)$ which on dimensional grounds excludes this quiver from the framed quotient quiver family in Table \ref{table:framed_ortho_quotient_quivers}.

\section{Rules for framed quotient quiver subtraction}
\label{sec:rules}
Having listed the framed orthosymplectic quotient quivers in Table \ref{table:framed_ortho_quotient_quivers}, it is now time to establish the set of rules for the \hyperref[sec:rules]{quotient quiver subtraction} algorithm. These share some features with both unitary \cite{Hanany:2023tvn} and unframed orthosymplectic \hyperref[sec:rules]{quotient quiver subtraction} \cite{Bennett:2024llh}, although importantly also contain differences. These comparisons will be touched on in Section \ref{subsec:comparison}.

It is imperative to reiterate that these rules will delineate the cases where a Coulomb branch global symmetry subgroup will be gauged with ``complete Higgsing". These cases are simple to check at the level of the Coulomb branch Hilbert series.

Firstly, a review of some terms that will be recurring throughout; the \textit{target quiver} is the quiver from which the quotient quiver is being subtracted, a \textit{long leg} is a chain of gauge nodes of the target quiver with no flavours on any gauge nodes, and a \textit{junction} is a gauge node of the target quiver which has gauge valency\footnote{The term `gauge valency' refers to the number of gauge nodes a particular node is connected to.} greater than $2$.

\begin{itemize}
\item A $G=\sorm(2n),\;\sorm(2n+1),\;\sprm(n)$ orthosymplectic framed quotient quiver must be aligned against a long leg starting with a balanced maximal chain of gauge nodes going up to at least $G^\vee$. The $G^\vee$ gauge node need not be balanced. The long leg from which the quotient quiver is being subtracted must not have any flavour nodes attached except on the gauge node on which the quotient quiver ends. The quotient quiver is permitted to go just one node past a junction.
\item The subtraction of gauge nodes proceeds by reducing the ranks of the gauge nodes in the target quiver by the rank of the gauge node of the quotient quiver aligned with it. The type of the algebra after subtraction changes for $D_n$ and $B_n$ algebras -- summarised below
\begin{equation}
    \begin{tikzpicture}
        \node (1) [gauger, label=below:{$D_n$}] at (0,0) {};
        \node (2) [gauger, label=below:{$D_m$}] at (1,0) {};
        \node (3) [gauger, label=below:{$B_{n-m}$}] at (2.5,0) {};
        \node at (0.5,0) {$-$};
        \node at (1.75,0) {$\Rightarrow$};
    \end{tikzpicture}
\end{equation}
\begin{equation}
    \begin{tikzpicture}
        \node (1) [gauger, label=below:{$B_n$}] at (0,0) {};
        \node (2) [gauger, label=below:{$B_m$}] at (1,0) {};
        \node (3) [gauger, label=below:{$D_{n-m}$}] at (2.5,0) {};
        \node at (0.5,0) {$-$};
        \node at (1.75,0) {$\Rightarrow$};
    \end{tikzpicture}
 \end{equation}
 \begin{equation}
    \begin{tikzpicture}
        \node (1) [gaugeb, label=below:{$C_n$}] at (0,0) {};
        \node (2) [gaugeb, label=below:{$C_m$}] at (1,0) {};
        \node (3) [gaugeb, label=below:{$C_{n-m}$}] at (2.5,0) {};
        \node at (0.5,0) {$-$};
        \node at (1.75,0) {$\Rightarrow$};
    \end{tikzpicture}
    \end{equation}

    Since the entire long leg of the target quiver from which the quotient quiver is being subtracted does not have flavours, situations where ``$B_{n}-D_{m}$" or vice-versa never occur -- otherwise the target quiver would be anomalous.  

    \item \textbf{Rebalancing Rule:} All gauge nodes are rebalanced using flavour groups.
    \item \textbf{Junction Rule:} If the quotient quiver goes one node past a junction, perform the subtraction for each end-point in turn. The result is the union of all of these possibilities.
\end{itemize}

\paragraph{Note} There is no known list of minimal degenerations for framed or unframed orthosymplectic quivers. Intersections of quivers are therefore found on a case-by-case basis.

\subsection{Comparison to previously known quotient quiver subtraction algorithms}
\label{subsec:comparison}
\epigraph{Quotient quiver subtraction doesn't repeat itself, but it often rhymes.}{\textit{Anonymous}}
The rules presented in Section \ref{sec:rules} bear several similarities and differences to those established for unitary and unframed orthosymplectic \hyperref[sec:rules]{quotient quiver subtraction}. As a review, the four quotient quivers found in \cite{Bennett:2024llh} for unframed orthosymplectic theories are reproduced in Table \ref{table:unframed_quotient_quivers}. Features shared by all three \hyperref[sec:rules]{quotient quiver subtraction} algorithms include the fact that subtraction occurs on maximal chains of gauge nodes only, as well as the presence of unions of cones resulting from the junction rule.

Interestingly, the framed orthosymplectic quotient quivers are of rank $\text{dim}(G)$, while the unframed quotient quivers in Table \ref{table:unframed_quotient_quivers} are of rank $\text{dim}(G)+1$. This reflects the fact that subtractions on unframed quivers are rebalanced using a $C_1$ gauge node while framed quivers are rebalanced with flavours. For unitary \hyperref[sec:rules]{quotient quiver subtraction}, rebalancing can be performed either with a single $\urm(1)$ gauge node or with flavours -- the two options result in equivalent Coulomb branches since the monopole formula converges for framed unitary quivers only.
\begin{table}[h!]
\ra{1.5}
    \centering
    \begin{tabular}{cc}
    \toprule
         $G$ & Unframed Quotient Quiver \\ \midrule
         $\surm(2)$ & 
         \raisebox{-0.5\height}{\begin{tikzpicture}
         \node[gauger, label=below:$D_1$] (d1l)[]{};
         \node[gaugeb, label=below:$C_1$] (c1l)[right=of d1l]{};
         \node[gauger, label=below:$D_1$] (d1r) [right=of c1l]{};
         \node[gaugeb, label=below:$C_1$] (c1r) [right=of d1r]{};
         \draw[-] (d1l)--(c1l)--(d1r)--(c1r);
        \end{tikzpicture}}
         \\ \midrule
         $\surm(3)$ & \raisebox{-0.5\height}{\begin{tikzpicture}
        \node[gaugeb, label=below:$C_1$] (c1rs) []{};
        \node[gauger, label=below:$D_2$] (d2rs) [left=of c1rs]{};
        \node[gaugeb, label=below:$C_2$] (c2s) [left=of d2rs]{};
        \node[gauger, label=below:$D_2$] (d2s) [left=of c2s]{};
        \node[gaugeb, label=below:$C_1$] (c1s) [left=of d2s]{};
        \node[gauger, label=below:$D_1$] (d1s) [left=of c1s]{};
        \draw[-] (c1rs)--(d2rs)--(c2s)--(d2s)--(c1s)--(d1s);
    \end{tikzpicture}} \\ \midrule
         $G_2$& \raisebox{-0.5\height}{\begin{tikzpicture}
        \node[gauger, label=below:$D_1$] (d1) []{};
        \node[gaugeb, label=below:$C_1$] (c1l) [right=of d1]{};
        \node[gauger, label=below:$D_2$] (d2l) [right=of c1l]{};
        \node[gaugeb, label=below:$C_2$] (c2) [right=of d2l]{};
        \node[gauger, label=below:$D_3$] (d3) [right=of c2]{};
        \node[gaugeb, label=below:$C_3$] (c3) [right=of d3]{};
        \node[gauger, label=below:$D_2$] (d2r) [right=of c3]{};
        \node[gaugeb, label=below:$C_1$] (c1r) [right=of d2r]{};
        \draw[-] (d1)--(c1l)--(d2l)--(c2)--(d3)--(c3)--(d2r)--(c1r);
    \end{tikzpicture}} \\
         \midrule
         $\sorm(7)$ & \raisebox{-0.5\height}{\begin{tikzpicture}
        \node[gauger, label=below:$D_1$] (d1) []{};
        \node[gaugeb, label=below:$C_1$] (c1l) [right=of d1]{};
        \node[gauger, label=below:$D_2$] (d2l) [right=of c1l]{};
        \node[gaugeb, label=below:$C_2$] (c2) [right=of d2l]{};
        \node[gauger, label=below:$D_3$] (d3l) [right=of c2]{};
        \node[gaugeb, label=below:$C_3$] (c3l) [right=of d3l]{};
        \node[gauger, label=below:$D_4$] (d4) [right=of c3l]{};
        \node[gaugeb, label=below:$C_3$] (c3r) [right=of d4]{};
        \node[gauger, label=below:$D_2$] (d2r) [right=of c3r]{};
        \node[gaugeb, label=below:$C_1$] (c1r) [right=of d2r]{};
        \draw[-] (d1)--(c1l)--(d2l)--(c2)--(d3l)--(c3l)--(d4)--(c3r)--(d2r)--(c1r);
    \end{tikzpicture}} \\
         \bottomrule
    \end{tabular}
    \caption{The set of quotient quivers for \emph{unframed} orthosymplectic quivers \cite{Bennett:2024llh}.}
    \label{table:unframed_quotient_quivers}
\end{table}

Another difference between the algorithms for framed and unframed orthosymplectic quivers consists of the fact that framed orthosymplectic \hyperref[sec:rules]{quotient quiver subtraction} changes $D$-type gauge nodes into $B$-type gauge nodes (and vice versa). This change in gauge algebra is vaguely reminiscent of those observed in Kraft-Procesi transitions in Hanany-Witten Type IIB brane systems with orientifold planes \cite{Cabrera:2016vvv,Cabrera:2017njm}, and is not observed in the unframed orthosymplectic \hyperref[sec:rules]{quotient quiver subtraction} algorithm.

\subsection{Weyl integration}
In order to verify the algorithm presented in Section \ref{sec:rules}, it is necessary to perform an independent check using Hilbert series methods.

Recall that a $G=\sorm(n),\;\sprm(n)$ \hyperref[sec:rules]{quotient quiver subtraction} gauges a subgroup $G$ of the Coulomb branch global symmetry, which realises a hyper-K\"ahler quotient by $G$ on the Coulomb branch. Furthermore, the rules in Section \ref{sec:rules} demonstrate how to gauge $G$ with complete Higgsing. This can be seen through the Hilbert series (HS) of the moduli space $\mathcal M$ with the Weyl integration formula \begin{equation}
    \mathrm{HS}\left[\mathcal M///G\right](t;\{x_i\})=\oint_{G}d\mu_G \mathrm{HS}\left[\mathcal M\right](t;\{x_i\},\{y_j\})\mathrm{PE}[-\chi_{\mathrm{Adj}}^G(\{y_j\})t^2]
\end{equation}
The original moduli space $\mathcal M$ has a global symmetry $G_{\mathcal M}$ of which a subgroup $G\subset G_{\mathcal M}$ is gauged leaving the commutant $C_{G_{\mathcal M}}(G)$. The notation in the above formula is as follows; $t$ is a fugacity for the R-charge, the $\{x_i\}$ are fugacities for $C_{G_{\mathcal M}}(G)$, and the $\{y_j\}$ are fugacities for $G$. The indices $i,j$ run from one to the rank of the respective groups.

The interpretation of the Weyl integration formula is to introduce additional F-terms which project out gauge invariant operators transforming in symmetric products of the adjoint representation of $G$, with a grading by the R-charge. This is denoted by the term $\mathrm{PE}[-\chi_{\mathrm{Adj}}^G(\{y_j\})t^2]$ where the $\chi$ stands for the character of the representation. The Haar measure $d\mu_G$ is the appropriate measure for the integral.

Although it is not known how to compute refined Coulomb branch Hilbert series for orthosymplectic quivers, there are libraries of these quivers for which the Coulomb branch is identified. Typically the Coulomb branch is a slice in a nilpotent cone with the quiver data corresponding to that for a nilpotent orbit \cite{Hanany:2016gbz,Hanany:2019tji}. Refined Hilbert series for some slices in nilpotent cones can also be computed using localisation techniques \cite{Hanany:2017ooe,Cabrera:2018ldc,Hanany:2019tji} -- as such, the Weyl integration formula can be applied to the Hilbert series of a given slice and its result checked against that arising from the \hyperref[sec:rules]{quotient quiver subtraction}.

\section{Simple examples}
\label{sec:simple_examples}
In this section, the \hyperref[sec:rules]{quotient quiver subtraction} algorithm is demonstrated on some framed orthosymplectic quivers whose moduli spaces of vacua are slices in nilpotent cones.
\subsection{$\overline{\mathcal O}^{\sorm(8)}_{(2^4)}///\sorm(3)$}
The orthosymplectic magnetic quiver for one of the nilpotent orbits of $\sorm(8)$ with very-even partition $(2^4)$, $\overline{\mathcal O}^{\sorm(8)}_{(2^4)}$, is given in the top of \Figref{fig:D4SO3Quot}. The action of an $\sorm(3)$ \hyperref[sec:rules]{quotient quiver subtraction} results in \Quiver{fig:D4SO3Quot}. The change in the algebra type from $D_2$ into $B_1$ should be noted in addition to the rebalancing of the $C_1$ gauge node with a $B_0$ flavour.

The Coulomb branch Hilbert series of \Quiver{fig:D4SO3Quot} is computed as \begin{equation}
    \hsC{fig:D4SO3Quot}=\frac{1 + 4 t^2 + 4 t^4 + t^6}{(1 - t^2)^6}
\end{equation}which identifies the moduli space as the $\overline{\mathcal O}^{\sorm(5)}_{(3,1^2)}$ \cite{Hanany:2016gbz}.

The action on the Coulomb branch is a hyper-Kähler quotient on $\overline{\mathcal O}^{\sorm(8)}_{(2^4)}$ by $\sorm(3)$ which is computed using the Weyl integration formula. This is performed using the following embedding of $\sorm(8)\hookleftarrow\sorm(3)\times \sorm(5)$ which decomposes the fundamental as \begin{equation}
    (\mu_1)_{\sorm(8)}\rightarrow (\mu_1)_{\sorm(5)}+\nu^2_{\sorm(3)}
\end{equation} where the $\mu_i$ and $\nu$ are highest weight fugacities of the respective groups.

The conclusion is that \begin{equation}
    \overline{\mathcal O}^{\sorm(8)}_{(2^4)}///\sorm(3)=\overline{\mathcal O}^{\sorm(5)}_{(3,1^2)}
\end{equation}

The Higgs branch Hilbert series of \Quiver{fig:D4SO3Quot} is evaluated as \begin{equation}
    \hsH{fig:D4SO3Quot}=\pe\left[[2]_{\sprm(1)}t^2+[1]_{\sprm(1)}t^3-t^8\right]
\end{equation}where the Dynkin label is shorthand for the given $\sprm(1)$ representation, which identifies the moduli space as $\mathcal S^{\sprm(2)}_{\mathcal N,(2,1^2)}$ as expected from Lusztig-Spaltenstein duality.

\begin{figure}[h!]
    \centering
    \begin{tikzpicture}
        \node[gauger, label=below:$D_1$] (d1l) at (0,0){};
        \node[gaugeb, label=below:$C_1$] (c1l) at (1,0){};
        \node[gauger, label=below:$D_2$] (d2) at (2,0){};
        \node[gaugeb, label=below:$C_1$] (c1r) at (3,0){};
        \node[gauger, label=below:$D_1$] (d1r) at (4,0){};
        \node[flavourb, label=above:$C_1$] (c1f) at (2,1){};
        \draw[-] (d1l)--(c1l)--(d2)--(c1r)--(d1r) (d2)--(c1f);

        \node[gauger, label=below:$D_1$] (d1ls) at (0,-1){};
        \node[gaugeb, label=below:$C_1$] (c1s) at (1,-1){};
        \node[gauger, label=below:$D_1$] (d1rs) at (2,-1){};

        \draw[-] (d1ls)--(c1s)--(d1rs);

        \node[] (minus) at (-1,-1){$-$};

        \node[gauger, label=below:$B_1$] (B1) at (2,-3){};
        \node[gaugeb, label=below:$C_1$] (C1) at (3,-3){};
        \node[gauger, label=below:$D_1$] (D1) at (4,-3){};
        \node[flavourb, label=left:$C_1$] (C1f) at (2,-2){};
        \node[flavourr, label=right:$B_0$] (B0f) at (3,-2){};

        \draw[-] (C1f)--(B1)--(C1)--(D1)  (C1)--(B0f);
    \end{tikzpicture}
    \caption{$\sorm(3)$ \hyperref[sec:rules]{quotient quiver subtraction} on the magnetic quiver for $\overline{\mathcal O}^{\sorm(8)}_{(2^4)}$ to produce \Quiver{fig:D4SO3Quot}.}
    \label{fig:D4SO3Quot}
\end{figure}

\subsection{$\overline{\mathcal O}^{\sorm(10)}_{(2^4,1^2)}///\sorm(3)$}
The orthosymplectic magnetic quiver for $\overline{\mathcal O}^{\sorm(10)}_{(2^4,1^2)}$ is given in the top of \Figref{fig:D5SO3Quot}. This is checked with unrefined Hilbert series computations which are not reproduced here. Performing an $\sorm(3)$ \hyperref[sec:rules]{quotient quiver subtraction} results in \Quiver{fig:D5SO3Quot}, as shown in \Figref{fig:D5SO3Quot}.

The Coulomb branch Hilbert series of \Quiver{fig:D5SO3Quot} is computed as \begin{equation}
    \hsC{fig:D5SO3Quot}=\frac{(1 + t^2) (1 + 6 t^2 + 21 t^4 + 28 t^6 + 21 t^8 + 6 t^{10} + 
   t^{12})}{(1 - t^2)^{14}}
\end{equation} which identifies the moduli space as the $\overline{\mathcal O}^{\sorm(7)}_{(3^2,1)}$ \cite{Hanany:2016gbz}.

The action on the Coulomb branch is a hyper-Kähler quotient on $\overline{\mathcal O}^{\sorm(10)}_{(2^4,1^2)}$ by $\sorm(3)$ which is computed using the Weyl integration formula. This involves the embedding $\sorm(10)\hookleftarrow\sorm(3)\times \sorm(7)$ which decomposes the fundamental as \begin{equation}
    (\mu_1)_{\sorm(10)}\rightarrow (\mu_1)_{\sorm(7)}+\nu^2_{\sorm(3)}
\end{equation} where the $\mu_i$ and $\nu$ are highest weight fugacities of the respective groups. The conclusion is that \begin{equation}
    \overline{\mathcal O}^{\sorm(10)}_{(2^4,1^2)}///\sorm(3)=\overline{\mathcal O}^{\sorm(7)}_{(3^2,1)}
\end{equation}

The Higgs branch Hilbert series of \Quiver{fig:D5SO3Quot} is evaluated as \begin{equation}
    \hsH{fig:D5SO3Quot}=\pe\left[[2]_{\sorm(3)}t^2+[4]_{\sorm(3)}t^4-t^8-t^{12}\right]
\end{equation} where the Dynkin label is shorthand for the given $\sorm(3)$ representation, which identifies the moduli space as $\mathcal S^{\sprm(3)}_{\mathcal N,(2^3)}$ as expected from Luzstig-Spaltenstein duality.
\begin{figure}[h!]
    \centering
    \begin{tikzpicture}
        \node[gauger, label=below:$D_1$] (d1l) at (0,0){};
        \node[gaugeb, label=below:$C_1$] (c1l) at (1,0){};
        \node[gauger, label=below:$D_2$] (d2l) at (2,0){};
        \node[gaugeb, label=below:$C_2$] (c2) at (3,0){};
        \node[gauger, label=below:$D_2$] (d2r) at (4,0){};
        \node[gaugeb, label=below:$C_1$] (c1r) at (5,0){};
        \node[gauger, label=below:$D_1$] (d1r) at (6,0){};
        \node[flavourr, label=above:$D_1$] (d1f) at (3,1){};
        \draw[-] (d1l)--(c1l)--(d2l)--(c2)--(d2r)--(c1r)--(d1r) (c2)--(d1f);

        \node[gauger, label=below:$D_1$] (d1ls) at (0,-1){};
        \node[gaugeb, label=below:$C_1$] (c1s) at (1,-1){};
        \node[gauger, label=below:$D_1$] (d1rs) at (2,-1){};

        \draw[-] (d1ls)--(c1s)--(d1rs);

        \node[] (minus) at (-1,-1){$-$};

        \node[flavourr, label=above:$B_1$] (B1f) at (3,-2){};
        \node[gaugeb, label=below:$C_2$] (C2) at (3,-3){};
        \node[gauger, label=below:$B_1$] (B1l) at (2,-3){};
        \node[gauger, label=below:$D_2$] (D2r) at (4,-3){};
        \node[gaugeb, label=below:$C_1$] (C1r) at (5,-3){};
        \node[gauger, label=below:$D_1$] (D1r) at (6,-3){};

        \draw[-] (B1l)--(C2)--(D2r)--(C1r)--(D1r) (C2)--(B1f);
    \end{tikzpicture}
    \caption{$\sorm(3)$ \hyperref[sec:rules]{quotient quiver subtraction} on the magnetic quiver for $\overline{\mathcal O}^{\sorm(10)}_{(2^4,1^2)}$ to produce \Quiver{fig:D5SO3Quot}.}
    \label{fig:D5SO3Quot}
\end{figure}

These results have a very simple realisation on the $3d$ mirror quiver as a gauging of an $\sorm(3)$ subgroup of the flavour symmetry. This is shown below \begin{equation}
    \begin{tikzpicture}
        \node[gaugeb, label=below:$C_2$] (c2) at (0,0){};
        \node[flavourr, label=above:$D_5$] (D5f) at (0,1){};
        \draw[-] (c2)--(D5f);
        \node[] (arrow) at (1,0.5){$\rightarrow$};
        \node[gaugeb, label=below:$C_2$] (C2) at (2,0){};
        \node[gauger, label=below:$B_1$] (B1) at (3,0){};
        \node[flavourr, label=above:$B_3$] (B3f) at (2,1){};
        \draw[-] (B1)--(C2)--(B3f);
    \end{tikzpicture}
\end{equation}

\begin{figure}[h!]
\centering
\begin{tikzpicture}
    \node (a) at (0,0){$\begin{tikzpicture}
         \node[gauger, label=below:$D_1$] (d1l) at (0,0){};
        \node[gaugeb, label=below:$C_1$] (c1l) at (1,0){};
        \node[gauger, label=below:$D_2$] (d2l) at (2,0){};
        \node[gaugeb, label=below:$C_2$] (c2) at (3,0){};
        \node[gauger, label=below:$D_2$] (d2r) at (4,0){};
        \node[gaugeb, label=below:$C_1$] (c1r) at (5,0){};
        \node[gauger, label=below:$D_1$] (d1r) at (6,0){};
        \node[flavourr, label=above:$D_1$] (d1f) at (3,1){};
        \draw[-] (d1l)--(c1l)--(d2l)--(c2)--(d2r)--(c1r)--(d1r) (c2)--(d1f);
    \end{tikzpicture}$};
    
    \node (b) at (9,0){$\begin{tikzpicture}
       \node[gaugeb, label=below:$C_2$] (c2) at (0,0){};
        \node[flavourr, label=above:$D_5$] (D5f) at (0,1){};
        \draw[-] (c2)--(D5f);\end{tikzpicture}$};

    \node (c) at (0,-5){$\begin{tikzpicture}
       \node[flavourr, label=above:$B_1$] (B1f) at (3,-2){};
        \node[gaugeb, label=below:$C_2$] (C2) at (3,-3){};
        \node[gauger, label=below:$B_1$] (B1l) at (2,-3){};
        \node[gauger, label=below:$D_2$] (D2r) at (4,-3){};
        \node[gaugeb, label=below:$C_1$] (C1r) at (5,-3){};
        \node[gauger, label=below:$D_1$] (D1r) at (6,-3){};

        \draw[-] (B1l)--(C2)--(D2r)--(C1r)--(D1r) (C2)--(B1f);
    \end{tikzpicture}$};

    \node (d) at (9,-5){$\begin{tikzpicture}
       \node[gaugeb, label=below:$C_2$] (C2) at (2,0){};
        \node[gauger, label=below:$B_1$] (B1) at (3,0){};
        \node[flavourr, label=above:$B_3$] (B3f) at (2,1){};
        \draw[-] (B1)--(C2)--(B3f);
    \end{tikzpicture}$};
    \draw[<->] (a)--(b)node[pos=0.5, above]{$3d$ Mirror Symmetry};
    \draw[<->] (c)--(d)node[pos=0.5, below]{$3d$ Mirror Symmetry};
    \draw[->] (a)--(c)node[pos=0.5, left]{$\sorm(3)$ \hyperref[sec:rules]{QQS}};
    \draw[->] (b)--(d)node[pos=0.5, right]{$\sorm(3)$ Flavour Gauge};
    \end{tikzpicture}
    \caption{Commutative diagram showing $\sorm(3)$ \hyperref[sec:rules]{quotient quiver subtraction} (QQS) down the left column and $\sorm(3)$ flavour symmetry gauging on the $3d$ mirror theory in the right column.}
    \label{fig:D5commute}
\end{figure}
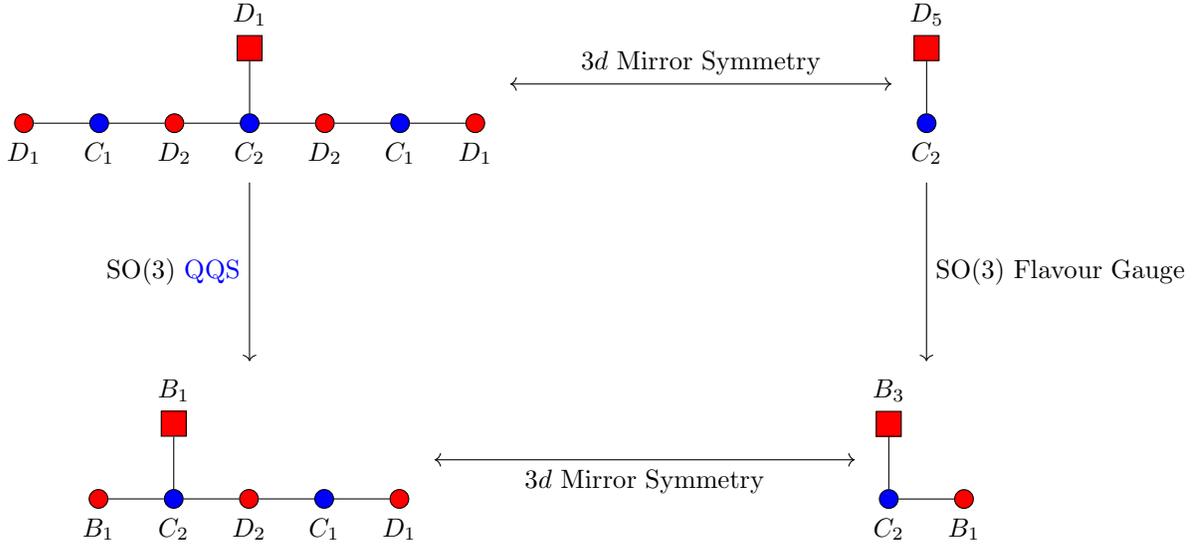

Once again \hyperref[sec:rules]{quotient quiver subtraction} and flavour symmetry gauging may be viewed as dual operations under $3d$ mirror symmetry, as summarised in \Figref{fig:D5commute}.

\section{Further examples}
\label{sec:examples}
Having warmed up with a few simple examples, \hyperref[sec:rules]{quotient quiver subtraction} is illustrated further with more non-trivial examples. 

The main library of quivers from which examples are chosen is \cite{Gaiotto:2008ak,Sperling:2021fcf}. The Coulomb branches of these quivers are quite peculiar in that they are products of two moduli spaces -- typically products of nilpotent orbit closures. This non-triviality makes for a robust test of \hyperref[sec:rules]{quotient quiver subtraction} since it may be checked independently with Weyl integration. The product of moduli spaces is achieved by introducing ON$^-$ planes to certain Type IIB brane systems.

For example, the one-parameter family of quivers \Quiver{fig:productD} in \Figref{fig:productD} has a Coulomb branch which is a product of nilpotent orbits \begin{equation}
    \Coul{fig:productD}=\overline{\mathcal O}^{D_{2k+1}}_{(2^{2k},1^2)}\times \overline{\mathcal O}^{D_{2k+1}}_{(2^{2k},1^2)}
\end{equation}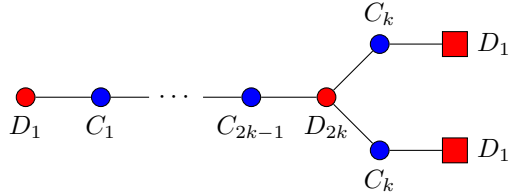
\begin{figure}[h!]
    \centering
    \begin{tikzpicture}
        \node[gauger, label=below:$D_1$] (d1l) at (0,0) {};
        \node[gaugeb, label=below:$C_1$] (c1l) at (1,0){};
        \node[] (cdots) at (2,0) {$\cdots$};
        \node[gaugeb, label=below:$C_{2k-1}$](ckm1) at (3,0){};
        \node[gauger, label=below:$D_{2k}$] (dk) at (4,0){};

        \node[gaugeb, label=above:$C_k$] (ckt) at ({4+cos(45)},{sin(45)}){};
        \node[gaugeb, label=below:$C_k$] (ckb) at ({4+cos(45)},{-sin(45)}){};

        \node[flavourr, label=right:$D_1$] (d1t) at ({5+cos(45)},{sin(45)}){};
        
        \node[flavourr, label=right:$D_1$] (d1b) at ({5+cos(45)},{-sin(45)}){};
        
        \draw[-] (d1l)--(c1l)--(cdots)--(ckm1)--(dk) (d1t)--(ckt)--(dk)--(ckb)--(d1b);     
    \end{tikzpicture}
    \caption{Example of a single quiver which has a Coulomb branch which is a product of moduli spaces.}
    \label{fig:productD}
\end{figure}
\subsection{$(\overline{min. D_3}\times\overline{min. D_3})///\sorm(3)$}
One of the product theories of \cite{Sperling:2021fcf} has Coulomb branch which is $\overline{min D_3}\times\overline{min. D_3}$, this theory was also studied in \cite{Gaiotto:2008ak}. This is the $k=1$ case of \Quiver{fig:productD} and is also shown at the top of \Figref{fig:minD3SqSO3}.

Performing $\sorm(3)$ \hyperref[sec:rules]{quotient quiver subtraction} results in the magnetic quiver \Quiver{fig:minD3SqSO3} shown in the bottom of \Figref{fig:minD3SqSO3}. Note that each $C_1$ gauge node in the bifurcation must be rebalanced with one half-hyper i.e. a $B_0$ flavour node. Since there is a $D_1$ flavour node already attached to each $C_1$ gauge node, the effect of rebalancing enhances the $D_1$ flavour to $B_1$.
\begin{figure}[h!]
    \centering
    \begin{tikzpicture}
        \node[gauger, label=below:$D_1$] (d1) at (0,0){};
        \node[gaugeb, label=below:$C_1$] (c1) at (1,0){};
        \node[gauger, label=below:$D_2$] (d2) at (2,0){};
        \node[gaugeb, label=above:$C_1$] (c1t) at ({2+cos(45)},{sin(45)}){};
        \node[gaugeb, label=below:$C_1$] (c1b) at ({2+cos(45)},{-sin(45)}){};
        \node[flavourr, label=right:$D_1$] (d1ft) at ({3+cos(45)},{sin(45)}){};
        \node[flavourr, label=right:$D_1$] (d1fb) at ({3+cos(45)},{-sin(45)}){};
        \draw[-] (d1)--(c1)--(d2)--(c1t)--(d1ft) (d2)--(c1b)--(d1fb);

        \node[gauger, label=below:$D_1$] (d1s) at (0,-2){};
        \node[gaugeb, label=below:$C_1$] (c1s) at (1,-2){};
        \node[gauger, label=below:$D_1$] (d1sr) at (2,-2){};
        \draw[-] (d1s)--(c1s)--(d1sr);
        \node[] (minus)at (-1,-2){$-$};

        \node[gauger, label=below:$B_1$] (B1) at (2,-4){};
        \node[gaugeb, label=above:$C_1$] (C1t) at ({2+cos(45)},{-4+sin(45)}){};
        \node[gaugeb, label=below:$C_1$] (C1b) at ({2+cos(45)},{-4-sin(45)}){};
        \node[flavourr, label=right:$B_1$] (D1ft) at ({3+cos(45)},{-4+sin(45)}){};
        \node[flavourr, label=right:$B_1$] (D1fb) at ({3+cos(45)},{-4-sin(45)}){};

        \draw[-] (D1fb)--(C1b)--(B1)--(C1t)--(D1ft);
        
        \end{tikzpicture}
       
    \caption{$\sorm(3)$ \hyperref[sec:rules]{quotient quiver subtraction} on the magnetic quiver for $\overline{min. D_3}\times \overline{min. D_3}$ to produce \Quiver{fig:minD3SqSO3}.}
     \label{fig:minD3SqSO3}
\end{figure}

The Coulomb branch Hilbert series of \Quiver{fig:minD3SqSO3} is computed as \begin{align}
    \hsC{fig:minD3SqSO3}&=\frac{1 + 3 t^2 + 11 t^4 + 10 t^6 + 11 t^8 + 
 3 t^{10} + t^{12}}{(1 - t^2)^3 (1 - t^4)^3}\\
 \PL\left[\hsC{fig:minD3SqSO3}\right]&=6 t^2 + 8 t^4 - 15 t^6 - 4 t^8 +O(t^{10})
\end{align}where the moduli space has $\sorm(3)\times\sorm(3)$ global symmetry but does not have any particular name.

The same Hilbert series is found from Weyl integration. Each factor of $\overline{min. D_3}$ has $\sorm(6)$ global symmetry which is branched to $\sorm(6)\hookleftarrow\sorm(3)\times\sorm(3)$ in the following way \begin{equation}
    (\mu_1)_{\sorm(6)}\rightarrow \mu^2_{\sorm(3)}+\nu^2_{\sorm(3)}
\end{equation}where $\mu_1,\;\mu,$ and $\nu$ are highest weight fugacities for $\sorm(6),\;\sorm(3),$ and $\sorm(3)$ respectively. The quotient is taken with respect to a diagonal $\sorm(3)$.

It is currently not known how to compute refined Hilbert series with the monopole formula from orthosymplectic quivers. However, from Weyl integration a refined Hilbert series may be computed. For brevity the highest weight generating is presented instead \begin{equation}
    \hwg\left[\mathcal C\left(\text{\Quiver{fig:minD3SqSO3}}\right)\right]=\pe\left[(\mu_1^2 +\mu_2^2) t^2 + (\mu_1^2 \mu_2^2+1) t^4 + \mu_1^2 \mu_2^2 t^6 - \mu_1^4 \mu_2^4 t^{12}\right]
\end{equation}where the $\mu_{1,2}$ are highest weight fugacities for $\sorm(3)\times \sorm(3)$.

The starting magnetic quiver for $\overline{min. D_3}\times \overline{min. D_3}$ has a $3d$ mirror which is given by the following product quiver \cite{Sperling:2021fcf}\begin{equation}
    \begin{tikzpicture}
        \node[gaugeb, label=below:$C_1$] (c1l) at (0,0){};
        \node[flavourr, label=above:$D_3$] (d3l) at (0,1){};
        \node[gaugeb, label=below:$C_1$] (c1r) at (2,0){};
        \node[flavourr, label=above:$D_3$] (d3r) at (2,1){};
        \node[] (times) at (1,0.5){$\times$};
        \draw[-] (c1l)--(d3l) (c1r)--(d3r);
    \end{tikzpicture}
    \label{fig:Sp1D3Sq}
\end{equation}Gauging a diagonal $\sorm(3)$ flavour symmetry of this quiver results in the following quiver \begin{equation}
    \begin{tikzpicture}
        \node[gaugeb, label=below:$C_1$] (c1l) at (0,0){};
        \node[gauger, label=below:$B_1$] (b1) at (1,0){};
        \node[gaugeb, label=below:$C_1$] (c1r) at (2,0){};
        \node[flavourr, label=left:$B_1$] (b1l) at (0,1){};
        \node[flavourr, label=right:$B_1$] (b1r) at (2,1){};
        \draw[-] (b1l)--(c1l)--(b1)--(c1r)--(b1r);
    \end{tikzpicture}
\end{equation}which is the same as \Quiver{fig:minD3SqSO3}.

As gauging a diagonal flavour symmetry in the electric theory is dual to gauging diagonal Coulomb branch global symmetry in the magnetic theory, the conclusion is that \Quiver{fig:minD3SqSO3} is self-dual under $3d$ mirror symmetry. This is easily confirmed with computation of the refined Higgs branch Hilbert series to find agreement. The Hilbert series will not be reproduced again.

There is another known $3d$ mirror to \Quiver{fig:minD3SqSO3} which is the following quiver \begin{equation}
    \begin{tikzpicture}
        \node[gauger, label=below:$D_1$] (d1l) at (0,0){};
        \node[gaugeb, label=below:$C_1$] (c1) at (1,0){};
        \node[gauger, label=below:$D_1$] (d1r) at (2,0){};
        \node[flavourr, label=above:$D_2$] (d2f) at (1,1){};
        \draw[-] (d1l)--(c1)--(d1r) (c1)--(d2f);
    \end{tikzpicture}
    \label{fig:minD3SqSO3Mirror2}
\end{equation} It is simple to check that $\Coul{fig:minD3SqSO3Mirror2}=\Coul{fig:minD3SqSO3}$ and $\Higgs{fig:minD3SqSO3Mirror2}=\Higgs{fig:minD3SqSO3}$. Therefore giving a pair of quivers whose Coulomb branch and Higgs branch are the same.

The fact that there are two $3d$ mirrors to \Quiver{fig:minD3SqSO3}; itself and \Quiver{fig:minD3SqSO3Mirror2} is not strange. It is important to note that the quiver descriptions of these theories are the IR effective field theories and that $3d$ mirror symmetry is an IR duality. The UV physics may be different and is most easily seen in the brane systems that give rise to these theories.

Consider quiver \Quiver{fig:minD3SqSO3} (equivalently \Quiver{fig:minD3SqSO3Mirror2}), the gauge nodes may be thought of as forming an $A_3$ or a $D_3$ Dynkin diagram. These two perspectives gives rise to two possible brane systems in Type IIB which can give this quiver as an IR theory. These are drawn in \Figref{fig:minD3SqSO3Brane} and \Figref{fig:minD3SqSO3Brane2} corresponding to the ``$A_3$'' and ``$D_3$'' perspectives respectively . The obvious distinction is the presence of $\mathrm{O5}^-$ and $\mathrm{ON}^0$ planes in \Figref{fig:minD3SqSO3Brane2} -- this is the realisation of the $A_3\simeq D_3$ isomorphism.

The gauge theories that arise as worldvolume theories of these brane systems flow to the same IR fixed point, but do not necessarily originate from the same theory in the UV. This is expected since the two brane systems in \Figref{fig:minD3SqSO3BraneBoth} are different and should give different UV completions of the IR theory.

If one starts with \Figref{fig:minD3SqSO3Brane} and determines the $S$-dual brane system, one finds that \Quiver{fig:minD3SqSO3Mirror2} is the $3d$ mirror to \Quiver{fig:minD3SqSO3}. However, the brane system in \Figref{fig:minD3SqSO3Brane2} is self-dual under $S$-duality and therefore one finds that \Quiver{fig:minD3SqSO3} is self-mirror.

\begin{figure}[h!]
    \centering
    \begin{subfigure}{0.45\textwidth}
    \begin{tikzpicture}
        \draw[-] (0,-1)--(0,1) (2,-1)--(2,1) (4,-1)--(4,1) (6,-1)--(6,1);
        \draw[-] (0,0.5)--(6,0.5) (0,-0.5)--(6,-0.5);
        \draw[-] (2,0)--(4,0);
        \node[D5] at (0.5,0){};
        \node[D5] at (1,0){};
        \node[D5] at (1.5,0){};
        \node[D5] at (4.5,0){};
        \node[D5] at (5,0){};
        \node[D5] at (5.5,0){};
        \draw[dashed] (2,0)--(1.5,0) (1,0)--(0.5,0);
        \draw[dotted] (1.5,0)--(1,0) (0.5,0)--(0,0);
        \draw[dashed] (4,0)--(4.5,0) (5,0)--(5.5,0);
        \draw[dotted] (4.5,0)--(5,0) (5.5,0)--(6,0);
        \node[] at (1,0.8){$1$};
        \node[] at (3,0.8){$1$};
        \node[] at (5,0.8){$1$};
    \end{tikzpicture}
    \caption{}
    \label{fig:minD3SqSO3Brane}
    \end{subfigure}
    \begin{subfigure}{0.45\textwidth}
        \begin{tikzpicture}
            \draw[-] (0,-1.5)--(0,1.5) (2,-1.5)--(2,1.5) (4,-1.5)--(4,1.5);
            \draw[dashed] (6,-1.5)--(6,1.5);
            \draw[-] (0,0.5)--(4,0.5) (0,-0.5)--(4,-0.5) (0,0)--(2,0);
            \node[D5] at (2.5,0){};
            \node[D5] at (3,0){};
            \node[D5] at (3.5,0){};
            \draw[dashed] (2,0)--(2.5,0) (3,0)--(3.5,0);
            \draw[dotted] (2.5,0)--(3,0) (3.5,0)--(4,0);
            \draw[-] (2,1)--(6,1)--(4,0.75);
            \draw[-] (2,-1)--(6,-1)--(4,-0.75);

            \node[circle, draw, fill=white, label=right:$\mathrm{ON}^{-}$] at (6,0) {};

            \node at (6,1.8){$\mathrm{O5}^-$};

            \node at (1,0.8){$1$};
            \node at (3,0.8){$1$};
            \node at (5,1.3){$1$};
            
        \end{tikzpicture}
        \caption{}
        \label{fig:minD3SqSO3Brane2}
    \end{subfigure}
    \caption{Two possible Type IIB brane systems giving rise to \Quiver{fig:minD3SqSO3} as an electric quiver.}
    \label{fig:minD3SqSO3BraneBoth}
\end{figure}

The \hyperref[sec:rules]{quotient quiver subtraction} shown in \Figref{fig:minD3SqSO3} starts with the $3d$ mirror theory of \Quiver{fig:Sp1D3Sq} which can be engineered in Type IIB using $\mathrm{ON}^-$ and $\mathrm{O5}^-$ planes as in \cite{Gaiotto:2008ak,Sperling:2021fcf}. This explains why the \hyperref[sec:rules]{quotient quiver subtraction} in \Figref{fig:minD3SqSO3} results in \Quiver{fig:minD3SqSO3} rather than \Quiver{fig:minD3SqSO3Mirror2}.
\FloatBarrier

\subsection{$\left(\overline{\mathcal O}^{D_5}_{(2^4,1^2)}\times\overline{\mathcal O}^{D_5}_{(2^4,1^2)}\right)///\sorm(5)$}
The magnetic quiver for $\overline{min. D_3}\times\overline{min. D_3}$ studied in the previous example is the $k=1$ member of the family of quivers \Quiver{fig:productD}. Another member of the family is a magnetic quiver for $\overline{\mathcal O}^{D_5}_{(2^4,1^2)}\times\overline{\mathcal O}^{D_5}_{(2^4,1^2)}$ which is the $k=2$ member of \Quiver{fig:productD} and is shown in the top of \Figref{fig:D5SqSO5}.

The $\sorm(5)$ orthosymplectic quotient quiver may be subtracted as shown in \Figref{fig:D5SqSO5} to produce \Quiver{fig:D5SqSO5}. In a similar fashion to the previous example, each $C_2$ gauge node in the bifurcation is rebalanced with a $B_0$ flavour which promotes the already attached $D_1$ flavour to $B_1$.
\begin{figure}[h!]
    \centering
    \begin{tikzpicture}
        \node[gauger, label=below:$D_1$] (d1) at (0,0){};
        \node[gaugeb, label=below:$C_1$] (c1) at (1,0){};
        \node[gauger, label=below:$D_2$] (d2) at (2,0){};
        \node[gaugeb, label=below:$C_2$] (c2) at (3,0){};
        \node[gauger, label=below:$D_3$] (d3) at (4,0){};
        \node[gaugeb, label=below:$C_3$] (c3) at (5,0){};
        \node[gauger, label=below:$D_4$] (d4) at (6,0){};
        \node[gaugeb, label=above:$C_2$] (c2t) at ({6+cos(45)},{sin(45)}){};
        \node[flavourr, label=right:$D_1$] (d1ft) at ({7+cos(45)},{sin(45)}){};
        \node[gaugeb, label=below:$C_2$] (c2b) at ({6+cos(45)},{-sin(45)}){};
        \node[flavourr, label=right:$D_1$] (d1fb) at ({7+cos(45)},{-sin(45)}){};

        \draw[-] (d1)--(c1)--(d2)--(c2)--(d3)--(c3)--(d4)--(c2t)--(d1ft) (d4)--(c2b)--(d1fb);

        \node[gauger, label=below:$D_1$] (d1ls) at (0,-2){};
        \node[gaugeb, label=below:$C_1$] (c1ls) at (1,-2){};
        \node[gauger, label=below:$D_2$] (d2ls) at (2,-2){};
        \node[gaugeb, label=below:$C_2$] (c2ls) at (3,-2){};
        \node[gauger, label=below:$D_2$] (d2rs) at (4,-2){};
        \node[gaugeb, label=below:$C_1$] (c1rs) at (5,-2){};
        \node[gauger, label=below:$D_1$] (d1rs) at (6,-2){};

        \draw[-] (d1ls)--(c1ls)--(d2ls)--(c2ls)--(d2rs)--(c1rs)--(d1rs);

        \node[] (minus) at (-1,-2){$-$};

        \node[gauger, label=below:$B_1$] (B1) at (4,-4){};
        \node[gaugeb, label=below:$C_2$] (C2) at (5,-4){};
        \node[gauger, label=below:$B_3$] (B3) at (6,-4){};
        \node[gaugeb, label=above:$C_2$] (C2t) at ({6+cos(45)},{-4+sin(45)}){};
        \node[flavourr, label=right:$B_1$] (B1ft) at ({7+cos(45)},{-4+sin(45)}){};
        \node[gaugeb, label=below:$C_2$] (C2b) at ({6+cos(45)},{-4-sin(45)}){};
        \node[flavourr, label=right:$B_1$] (B1fb) at ({7+cos(45)},{-4-sin(45)}){};

        \draw[-] (B1)--(C2)--(B3)--(C2t)--(B1ft) (B3)--(C2b)--(B1fb);
    \end{tikzpicture}
    \caption{$\sorm(5)$ \hyperref[sec:rules]{quotient quiver subtraction} on the magnetic quiver for $\overline{O}^{D_5}_{(2^4,1^2)}\times\overline{O}^{D_5}_{(2^4,1^2)}$ to produce \Quiver{fig:D5SqSO5}.}
    \label{fig:D5SqSO5}
\end{figure}

The Coulomb branch Hilbert series is evaluated as \begin{align}
    \hsC{fig:D5SqSO5}&=\frac{\left(\begin{aligned}1 &+ 10 t^2 + 69 t^4 + 335 t^6 + 
   1300 t^8 + 3946 t^{10} + 9717 t^{12} \\&+ 19212 t^{14} + 31215 t^{16} + 
   41457 t^{18} + 45700 t^{20} + \cdots + 
   t^{40}\end{aligned}\right)}{(1-t^2)^{10}(1-t^4)^{10}}\\\PL\left[\hsC{fig:D5SqSO5}\right]&=20 t^2 + 24 t^4 - 25 t^6 - 40 t^8 - 61 t^{10}+O(t^{12})
\end{align}where the moduli space has $\sorm(5)\times\sorm(5)$ global symmetry but no particular name.

The explicit hyper-Kähler quotient construction of $\left(\overline{\mathcal O}^{D_5}_{(2^4,1^2)}\times\overline{\mathcal O}^{D_5}_{(2^4,1^2)}\right)///\sorm(5)$ is computed with Weyl integration. This is done using the following embedding of $\sorm(10)\hookleftarrow\sorm(5)\times\sorm(5)$ which decomposes the vector as \begin{equation}
    (\mu_1)_{\sorm(10)}\rightarrow (\mu_1)_{\sorm(5)}+(\nu_1)_{\sorm(5)}
\end{equation} and choosing a diagonal $\sorm(5)$ factor from each $\sorm(10)$.

The Higgs branch Hilbert series may also be computed as \begin{align}
    \hsH{fig:D5SqSO5}&=\frac{(1 + 6 t^4 + 6 t^6 + 26 t^8 + 15 t^{10} + 76 t^{12} + 30 t^{14} + 
 107 t^{16} + 50 t^{18} + \cdots + t^{36})}{(1 - t^2)^6 (1 - t^4)^3 (1 - t^8)^3}\\\PL\left[\hsH{fig:D5SqSO5}\right]&=6 t^2 + 9 t^4 + 6 t^6 + 8 t^8 - 21 t^{10} - 31 t^{12} +47 t^{16} +O(t^{18})
\end{align} The moduli space has no particular name but does have a global symmetry of $\sorm(3)\times \sorm(3)$ and is of dimension six as expected.

The magnetic quiver for $\overline{\mathcal O}^{D_5}_{(2^4,1^2)}\times\overline{\mathcal O}^{D_5}_{(2^4,1^2)}$ has a $3d$ mirror which is the product quiver \begin{equation}
    \begin{tikzpicture}
        \node[gaugeb, label=below:$C_2$] (c2l) at (0,0){};
        \node[flavourr, label=above:$D_5$] (d5l) at (0,1){};
        \node[gaugeb, label=below:$C_2$] (c2r) at (2,0){};
        \node[flavourr, label=above:$D_5$] (d5r) at (2,1){};
        \node[] (times) at (1,0.5){$\times$};
        \draw[-] (c2l)--(d5l) (c2r)--(d5r);
    \end{tikzpicture}\label{quiv:D5SqMirror}
\end{equation}Gauging a diagonal $\sorm(5)$ flavour symmetry of this quiver results in the following quiver \begin{equation}
    \begin{tikzpicture}
        \node[gaugeb, label=below:$C_2$] (c2l) at (0,0){};
        \node[gauger, label=below:$B_2$] (b2) at (1,0){};
        \node[gaugeb, label=below:$C_2$] (c2r) at (2,0){};
        \node[flavourr, label=left:$B_2$] (b2l) at (0,1){};
        \node[flavourr, label=right:$B_2$] (b2r) at (2,1){};
        \draw[-] (b2l)--(c2l)--(b2)--(c2r)--(b2r);
    \end{tikzpicture}
    \label{fig:D5SqSO5Mirror}
\end{equation}which is conjectured to be a $3d$ mirror of \Quiver{fig:D5SqSO5}. It is simple to verify through Hilbert series computations that $\Coul{fig:D5SqSO5Mirror}=\Higgs{fig:D5SqSO5}$ and $\Higgs{fig:D5SqSO5Mirror}=\Coul{fig:D5SqSO5}$.

From a brane construction there is a known $3d$ mirror to \Quiver{fig:D5SqSO5Mirror} which is \begin{equation}
    \begin{tikzpicture}
        \node[gauger, label=below:$D_1$] (d1l) at (0,0){};
        \node[gaugeb, label=below:$C_1$] (c1l) at (1,0){};
        \node[gauger, label=below:$D_2$] (d2l) at (2,0){};
        \node[gaugeb, label=below:$C_2$] (c2) at (3,0){};
        \node[gauger, label=below:$D_2$] (d2r) at (4,0){};
        \node[gaugeb, label=below:$C_1$] (c1r) at (5,0){};
        \node[gauger, label=below:$D_1$] (d1r) at (6,0){};
        \node[flavourr, label=above:$D_2$] (d2f) at (3,1){};

        \draw[-] (d1l)--(c1l)--(d2l)--(c2)--(d2r)--(c1r)--(d1r) (c2)--(d2f);
        
        \end{tikzpicture}
        \label{fig:D5SqSO5Mirror2}
\end{equation}This means that \Quiver{fig:D5SqSO5Mirror} and \Quiver{fig:D5SqSO5Mirror2} have the same Coulomb branch and the same Higgs branch. Indeed it is straightforward to verify through Hilbert series computations that $\Coul{fig:D5SqSO5}=\Coul{fig:D5SqSO5Mirror2}$ and $\Higgs{fig:D5SqSO5}=\Higgs{fig:D5SqSO5Mirror2}$.

Similar to the last example there are two possible Type IIB brane systems which give rise to \Quiver{fig:D5SqSO5Mirror}. These are drawn in \Figref{fig:D5SqSO5BraneBoth} where the distinction between the two brane systems is in the presence of an ON$^{-}$ or not.

\begin{figure}[h!]
    \centering
    \begin{subfigure}{0.45\textwidth}
    \begin{tikzpicture}
        \draw[-] (0,-1)--(0,1) (2,-1)--(2,1) (4,-1)--(4,1) (6,-1)--(6,1);
        \draw[-] (0,0.5)--(6,0.5) (0,-0.5)--(6,-0.5);
        \draw[-] (2,0)--(4,0);
        \node[D5] at ({2/6},0){};
        \node[D5] at ({4/6},0){};
        \node[D5] at ({6/6},0){};
        \node[D5] at ({8/6},0){};
        \node[D5] at ({10/6},0){};

        \node[D5] at ({4+2/6},0){};
        \node[D5] at ({4+4/6},0){};
        \node[D5] at ({4+6/6},0){};
        \node[D5] at ({4+8/6},0){};
        \node[D5] at ({4+10/6},0){};

        \draw[dashed] (2,0)--({10/6},0) ({8/6},0)--({6/6},0) ({4/6},0)--({2/6},0);
        \draw[dotted] ({10/6},0)--({8/6},0) ({6/6},0)--({4/6},0) ({2/6},0)--(0,0);
        \draw[dashed] (4,0)--({4+2/6},0) ({4+4/6},0)--({4+6/6},0) ({4+8/6},0)--({4+10/6},0);
        \draw[dotted] ({4+2/6},0)--({4+4/6},0) ({4+6/6},0)--({4+8/6},0) ({4+10/6},0)--({4+12/6},0);
        \node[] at (1,0.8){$2$};
        \node[] at (3,0.8){$2$};
        \node[] at (5,0.8){$2$};
    \end{tikzpicture}
    \caption{}
    \label{fig:D5SqSO5Brane}
    \end{subfigure}
    \begin{subfigure}{0.45\textwidth}
        \begin{tikzpicture}
            \draw[-] (0,-1.5)--(0,1.5) (2,-1.5)--(2,1.5) (4,-1.5)--(4,1.5);
            \draw[dashed] (6,-1.5)--(6,1.5);
            \draw[-] (0,0.5)--(4,0.5) (0,-0.5)--(4,-0.5) (0,0)--(2,0);
            \node[D5] at ({2+2/6},0){};
            \node[D5] at ({2+4/6},0){};
            \node[D5] at ({2+6/6},0){};
            \node[D5] at ({2+8/6},0){};
            \node[D5] at ({2+10/6},0){};
            \draw[dashed] (2,0)--({2+2/6},0) ({2+4/6},0)--({2+6/6},0) ({2+8/6},0)--({2+10/6},0);
            \draw[dotted] ({2+2/6},0)--({2+4/6},0) ({2+6/6},0)--({2+8/6},0) ({2+10/6},0)--({2+12/6},0);
            \draw[-] (2,1)--(6,1)--(4,0.75);
            \draw[-] (2,-1)--(6,-1)--(4,-0.75);

            \node[circle, draw, fill=white, label=right:$\mathrm{ON}^{-}$] at (6,0) {};

            \node at (6,1.8){$\mathrm{O5}^-$};

            \node at (1,0.8){$2$};
            \node at (3,0.8){$2$};
            \node at (5,1.3){$2$};
            
        \end{tikzpicture}
        \caption{}
        \label{fig:D5SqSO5Brane2}
    \end{subfigure}
    \caption{Two possible Type IIB brane systems giving rise to the quiver \Quiver{fig:D5SqSO5Mirror}.}
    \label{fig:D5SqSO5BraneBoth}
\end{figure}

Starting from the brane system in \Figref{fig:D5SqSO5Brane}, the $S$-dual brane system gives rise to the quiver \Quiver{fig:D5SqSO5Mirror}. Starting from the brane system in \Figref{fig:D5SqSO5Brane2}, the $S$-dual brane system gives rise to the quiver \Quiver{fig:D5SqSO5}.

The brane realisation of the product theory \Quiver{quiv:D5SqMirror} given in \cite{Sperling:2021fcf} uses a construction with ON$^-$ planes and hence the $\sorm(5)$ \hyperref[sec:rules]{quotient quiver subtraction} shown in \Figref{fig:D5SqSO5} gives rise to \Quiver{fig:D5SqSO5} instead of \Quiver{fig:D5SqSO5Mirror2}.

The interpretation is that the two brane systems shown in \Figref{fig:D5SqSO5BraneBoth} give rise to two different gauge theories that flow to the same IR fixed point and hence have the same moduli spaces of vacua in the IR, but have different UV completions.
\subsection{$\left(\overline{\mathcal O}^{D_5}_{(2^4,1^2)}\times\overline{\mathcal O}^{D_5}_{(2^4,1^2)}\right)///\sorm(4)$}
The very same magnetic quiver for $\overline{\mathcal O}^{D_5}_{(2^4,1^2)}\times\overline{\mathcal O}^{D_5}_{(2^4,1^2)}$ admits a subtraction of the $\sorm(4)$ quotient quiver. This is shown in \Figref{fig:D5SqSO4} to produce \Quiver{fig:D5SqSO4}. The rebalancing requires that the $C_1$ gauge node of \Quiver{fig:D5SqSO4} is rebalanced with a $B_0$ flavour.
\begin{figure}[h!]
    \centering
    \begin{tikzpicture}
        \node[gauger, label=below:$D_1$] (d1) at (0,0){};
        \node[gaugeb, label=below:$C_1$] (c1) at (1,0){};
        \node[gauger, label=below:$D_2$] (d2) at (2,0){};
        \node[gaugeb, label=below:$C_2$] (c2) at (3,0){};
        \node[gauger, label=below:$D_3$] (d3) at (4,0){};
        \node[gaugeb, label=below:$C_3$] (c3) at (5,0){};
        \node[gauger, label=below:$D_4$] (d4) at (6,0){};
        \node[gaugeb, label=above:$C_2$] (c2t) at ({6+cos(45)},{sin(45)}){};
        \node[flavourr, label=right:$D_1$] (d1ft) at ({7+cos(45)},{sin(45)}){};
        \node[gaugeb, label=below:$C_2$] (c2b) at ({6+cos(45)},{-sin(45)}){};
        \node[flavourr, label=right:$D_1$] (d1fb) at ({7+cos(45)},{-sin(45)}){};

        \draw[-] (d1)--(c1)--(d2)--(c2)--(d3)--(c3)--(d4)--(c2t)--(d1ft) (d4)--(c2b)--(d1fb);

        \node[gauger, label=below:$D_1$] (d1ls) at (0,-2){};
        \node[gaugeb, label=below:$C_1$] (c1ls) at (1,-2){};
        \node[gauger, label=below:$D_2$] (d2ls) at (2,-2){};
        \node[gaugeb, label=below:$C_1$] (c1rs) at (3,-2){};
        \node[gauger, label=below:$D_1$] (d1rs) at (4,-2){};
        
        \draw[-] (d1ls)--(c1ls)--(d2ls)--(c1rs)--(d1rs);

        \node[] (minus) at (-1,-2){$-$};

        \node[gaugeb, label=below:$C_1$] (C1) at (3,-4){};
        \node[gauger, label=below:$B_2$] (B2) at (4,-4){};
        \node[gaugeb, label=below:$C_3$] (C3) at (5,-4){};
        \node[gauger, label=below:$D_4$] (D4) at (6,-4){};
        \node[gaugeb, label=above:$C_2$] (C2t) at ({6+cos(45)},{-4+sin(45)}){};
        \node[flavourr, label=right:$D_1$] (D1ft) at ({7+cos(45)},{-4+sin(45)}){};
        \node[gaugeb, label=below:$C_2$] (C2b) at ({6+cos(45)},{-4-sin(45)}){};
        \node[flavourr, label=right:$D_1$] (D1fb) at ({7+cos(45)},{-4-sin(45)}){};
        \node[flavourr, label=left:$B_0$] (B0f) at (3,-3){};
        \node[flavourr, label=right:$B_0$] (B0f2) at (5,-3){};

        \draw[-] (C1)--(B2)--(C3)--(D4)--(C2t)--(D1ft) (D4)--(C2b)--(D1fb) (C1)--(B0f) (B0f2)--(C3);
    \end{tikzpicture}
    \caption{$\sorm(4)$ \hyperref[sec:rules]{quotient quiver subtraction} on the magnetic quiver for $\overline{\mathcal O}^{D_5}_{(2^4,1^2)}\times\overline{\mathcal O}^{D_5}_{(2^4,1^2)}$ to produce \Quiver{fig:D5SqSO4}.}
    \label{fig:D5SqSO4}
\end{figure}

The Coulomb branch Hilbert series of \Quiver{fig:D5SqSO4} is computed as \begin{align}
    \hsC{fig:D5SqSO4}&=\frac{\left(\begin{aligned}1 &+ 16 t^2 + 158 t^4 + 1166 t^6 + 
   7127 t^8 + 35858 t^{10} + 149380 t^{12} \\&+ 515052 t^{14} + 1481149 t^{16} + 3572884 t^{18} + 7280258 t^{20} + 12591678 t^{22} \\&+ 18569651 t^{24} + 
   23413746 t^{26} + 25291592 t^{28} + \cdots + t^{56}\end{aligned}\right)}{(1 - t^2)^{14} (1 - t^4)^{14}} \\\PL\left[\hsC{fig:D5SqSO4}\right]&=30 t^2 + 36 t^4 - 2 t^6+38t^8+O(t^{10})
\end{align}where the moduli space has $\sorm(6)\times\sorm(6)$ global symmetry but does not have any particular name.

The explicit hyper-Kähler quotient is computed with Weyl integration using the following embedding of $\sorm(10)\hookleftarrow\sorm(6)\times\sorm(4)$ which decomposes the vector as \begin{equation}
    (\mu_1)_{\sorm(10)}\rightarrow (\mu_1)_{\sorm(6)}+(\nu_1)_{\sorm(4)}
\end{equation}The Weyl integration is performed w.r.t a diagonal $\sorm(4)$ in each factor of $\sorm(10)$.
The Higgs branch Hilbert series is computationally challenging to find. Now consider the action of gauging a diagonal $\sorm(4)$ flavour symmetry of \Quiver{quiv:D5SqMirror} to produce the following quiver \begin{equation}
    \begin{tikzpicture}
         \node[gaugeb, label=below:$C_2$] (c2l) at (0,0){};
        \node[gauger, label=below:$D_2$] (d2) at (1,0){};
        \node[gaugeb, label=below:$C_2$] (c2r) at (2,0){};
        \node[flavourr, label=left:$D_3$] (d3l) at (0,1){};
        \node[flavourr, label=right:$D_3$] (d3r) at (2,1){};
        \draw[-] (d3l)--(c2l)--(d2)--(c2r)--(d3r);
    \end{tikzpicture}\label{quiv:D5SqSO4Mirror}
\end{equation}This quiver is conjectured to be the $3d$ mirror of \Quiver{fig:D5SqSO4}. It is simple to check with Hilbert series that $\Higgs{quiv:D5SqSO4Mirror}=\Coul{fig:D5SqSO4}$. It is also straightforward to compute the Coulomb branch as 
    \begin{align}
    \hsC{quiv:D5SqSO4Mirror}&=\frac{(1 - t^{12}) \left(1+ 5 t^4 + 6 t^6 + 22 t^8 + 30 t^{10} + 69 t^{12} + 
   90 t^{14} + 141 t^{16} + 146 t^{18} + 172 t^{20} + \cdots +t^{40}\right)}{(1 - t^2)^2  (1 - t^4)^5 (1 - t^6)^4 (1 - t^8)^2}\\
    \PL\left[\hsC{quiv:D5SqSO4Mirror}\right]&=2 t^2 + 10 t^4 + 10 t^6 + 9 t^8 - 23 t^{12} - 42 t^{14} - 57 t^{16} + O(t^{18})
\end{align}
where the moduli space has a $\urm(1)\times\urm(1)$ global symmetry but does not have any particular name. It is conjectured that $\Coul{quiv:D5SqSO4Mirror}=\Higgs{fig:D5SqSO4}$.

\begin{figure}[h!]
    \centering
    \begin{subfigure}{0.45\textwidth}
    \begin{tikzpicture}
        \draw[-] (0,-1)--(0,1) (2,-1)--(2,1) (4,-1)--(4,1) (6,-1)--(6,1);
        \draw[-] (0,0.5)--(6,0.5) (0,-0.5)--(6,-0.5);
        \node[D5] at ({2/7},0){};
        \node[D5] at ({4/7},0){};
        \node[D5] at ({6/7},0){};
        \node[D5] at ({8/7},0){};
        \node[D5] at ({10/7},0){};
        \node[D5] at ({12/7},0){};

        \node[D5] at ({4+2/7},0){};
        \node[D5] at ({4+4/7},0){};
        \node[D5] at ({4+6/7},0){};
        \node[D5] at ({4+8/7},0){};
        \node[D5] at ({4+10/7},0){};
        \node[D5] at ({4+12/7},0){};

        \draw[-] (2,0)--({12/7},0) ({10/7},0)--({8/7},0) ({6/7},0)--({4/7},0) ({2/7},0)--(0,0);
        \draw[dotted] ({12/7},0)--({10/7},0) ({8/7},0)--({6/7},0) ({4/7},0)--({2/7},0);

        \draw[-] (4,0)--({4+2/7},0) ({4+4/7},0)--({4+6/7},0) ({4+8/7},0)--({4+10/7},0) ({4+12/7},0)--(6,0);
        \draw[dotted] ({4+2/7},0)--({4+4/7},0) ({4+6/7},0)--({4+8/7},0) ({4+10/7},0)--({4+12/7},0);
        \node[] at (1,0.8){$2$};
        \node[] at (3,0.8){$2$};
        \node[] at (5,0.8){$2$};
    \end{tikzpicture}
    \caption{}
    \label{fig:D5SqSO4Brane}
    \end{subfigure}
    \begin{subfigure}{0.45\textwidth}
        \begin{tikzpicture}
            \draw[-] (0,-1.5)--(0,1.5) (2,-1.5)--(2,1.5) (4,-1.5)--(4,1.5);
            \draw[dashed] (6,-1.5)--(6,1.5);
            \draw[-] (0,0.5)--(4,0.5) (0,-0.5)--(4,-0.5);
            \node[D5] at ({2+2/7},0){};
            \node[D5] at ({2+4/7},0){};
            \node[D5] at ({2+6/7},0){};
            \node[D5] at ({2+8/7},0){};
            \node[D5] at ({2+10/7},0){};
            \node[D5] at ({2+12/7},0){};

            \draw[-] (2,1)--(6,1)--(4,0.75);
            \draw[-] (2,-1)--(6,-1)--(4,-0.75);

             \draw[-] (2,0)--({2+2/7},0) ({2+4/7},0)--({2+6/7},0) ({2+8/7},0)--({2+10/7},0) ({2+12/7},0)--(4,0);
            \draw[dotted] ({2+2/7},0)--({2+4/7},0) ({2+6/7},0)--({2+8/7},0) ({2+10/7},0)--({2+12/7},0);

            \node[circle, draw, fill=white, label=right:$\mathrm{ON}^{-}$] at (6,0) {};

            \node at (6,1.8){$\mathrm{O5}^-$};

            \node at (1,0.8){$2$};
            \node at (3,0.8){$2$};
            \node at (5,1.3){$2$};
            
        \end{tikzpicture}
        \caption{}
        \label{fig:D5SqSO4Brane2}
    \end{subfigure}
    \caption{Two Type IIB brane systems.}
    \label{fig:D5SqSO4BraneBoth}
\end{figure}
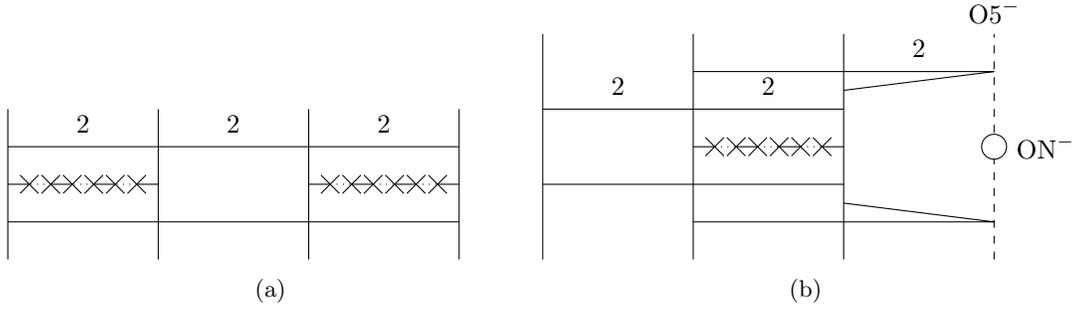

If the central $D_2$ gauge node in \Quiver{quiv:D5SqSO4Mirror} was $\orm(4)$ rather than $\sorm(4)$ then one can associate the brane system in \Figref{fig:D5SqSO4Brane} to it \cite{Bennett:2025zor}, but reiterate that this brane system does not give rise to \Quiver{quiv:D5SqSO4Mirror}.

However, starting first with the brane system in \Figref{fig:D5SqSO4Brane}, one suggests the following mirror to \Quiver{quiv:D5SqSO4Mirror} which is \Quiver{quiv:D5SqSO4Mirror2} where the central $D_2$ node is $\sorm(4)$ and not $\orm(4)$ as would be expected from such a brane construction.  
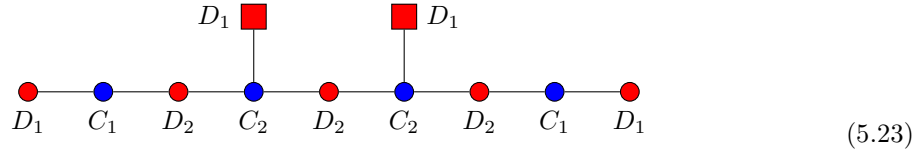
\begin{equation}
    \begin{tikzpicture}
        \node[gauger, label=below:$D_1$] (d1l) at (0,0){};
        \node[gaugeb, label=below:$C_1$] (c1l) at (1,0){};
        \node[gauger, label=below:$D_2$](d2l) at (2,0){};
        \node[gaugeb, label=below:$C_2$] (c2l) at (3,0){};
        \node[gauger, label=below:$D_2$] (d2m) at (4,0){};
        \node[gaugeb, label=below:$C_2$] (c2r) at (5,0){};
        \node[gauger, label=below:$D_2$] (d2r) at (6,0){};
        \node[gaugeb, label=below:$C_1$] (c1r) at (7,0){};
        \node[gauger, label=below:$D_1$] (d1r) at (8,0){};
        \node[flavourr, label=left:$D_1$] (d1fl) at (3,1){};
        \node[flavourr, label=right:$D_1$] (d1fr) at (5,1){};
        
        \draw[-] (d1l)--(c1l)--(d2l)--(c2l)--(d2m)--(c2r)--(d2r)--(c1r)--(d1r) (c2l)--(d1fl) (c2r)--(d1fr);
    \end{tikzpicture}\label{quiv:D5SqSO4Mirror2}
\end{equation}Indeed this magnetic quiver \Quiver{quiv:D5SqSO4Mirror2} is $3d$ mirror to \Quiver{quiv:D5SqSO4Mirror} and can be checked by computation with Hilbert series of its Coulomb and Higgs branch. 

If one starts with \Figref{fig:D5SqSO4Brane2}, which is a genuine brane system for \Quiver{quiv:D5SqSO4Mirror}, the corresponding mirror theory is \Quiver{fig:D5SqSO4}.

Therefore \Quiver{fig:D5SqSO4} and \Quiver{quiv:D5SqSO4Mirror2} is an example of another pair of quivers which have the same Coulomb branch and the same Higgs branch.
\section{Testing the junction rule}
\label{sec:testing_junction_rule}
A non-trivial and defining feature of the \hyperref[sec:rules]{quotient quiver subtraction} algorithm is in identifying unions of cones with all possible alignments of the quotient quiver corresponding to each cone. As the $\sorm(n)$ quotient quiver (for both $n$ even and odd) has an odd number of gauge nodes, unions of cones appear when it is subtracted from a leg with an even number of gauge nodes and there is a bifurcation in the quiver. The examples in \cite{Sperling:2021fcf} studied in the previous section have a bifurcation but the long leg has an odd number of gauge nodes meaning that unions of cones cannot appear. There are no other known examples in the literature which satisfy the criteria of having a long leg with an even number of gauge nodes, a bifurcation, and also identified moduli spaces.

In order to test the junction rule, a set of quivers must be constructed which have (partially) refined Hilbert series'. In the monopole formula, it is unknown how to refine Coulomb branch Hilbert series' of orthosymplectic quivers. Fortunately expressing the Coulomb branch Hilbert series as Hall-Littlewood polynomials \cite{Gadde:2011uv,Hanany:2015hxa} allows for the gluing of different legs with some refinement of the Coulomb branch global symmetry. In some cases it is possible to obtain fully refined Coulomb branch Hilbert series' for orthosymplectic quivers. It is then possible to test the junction rule on these quivers by comparing with the hyper-Kähler quotient.

Although possibly contrived, the following set of examples aim to illustrate the junction rule feature of this algorithm.

The Hilbert series for a union of cones (in this case Coulomb branches) takes a very simple form as a signed sum of intersections. Explicitly, for the union of a set of $n$ cones $\mathcal M_{1,2,\cdots,n}$, the Hilbert series is given as \begin{align}
    \hs\left[\mathcal M_1\cup\cdots\cup\mathcal M_n\right]=\sum_{r=1}^n(-1)^{r-1}\sum_{\substack{\text{r-tuples}\;(i_1,\cdots,i_r)\\\subset (1,2,\cdots,n)}}\hs\left[\mathcal M_{i_1}\cap\cdots\cap\mathcal M_{i_r}\right]
\end{align}

The intersection of $r$ moduli spaces may be found through $r$ repeated Kraft-Procesi transitions on each moduli space. On the Coulomb branch the Kraft-Procesi transitions may be realised through a different quiver subtraction algorithm \cite{Cabrera:2018ann,Bourget:2022tmw}. It is important to note that these algorithms have not been systematically extended to orthosymplectic quivers yet. The examples that are studied here are simple enough to naïvely extend those rules.

In the study of theories with eight supercharges, the Higgs branch of electric theories with one $\sprm(n)$ or $\surm(n)$ gauge node which has too few flavours will be a union of cones \cite{Ferlito:2016grh,Bourget:2023cgs}. In the magnetic theory, the union of cones is seen as \hyperref[sec:rules]{quotient quiver subtraction} with the junction rule \cite{Hanany:2023tvn,Bennett:2024llh}.
\paragraph{Example 1}
Consider the following quiver 
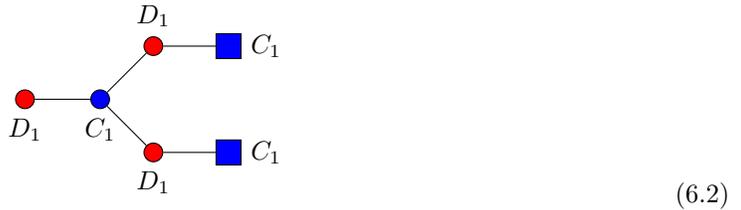
\begin{equation}
    \begin{tikzpicture}
        \node[gauger, label=below:$D_1$] (d1l) at (0,0){};
        \node[gaugeb, label=below:$C_1$] (c1l) at (1,0){};
        \node[gauger, label=above:$D_1$] (d1t) at ({1+cos(45)},{sin(45)}){};
        \node[gauger, label=below:$D_1$] (d1b) at ({1+cos(45)},{-sin(45)}){};
        \node[flavourb, label=right:$C_1$] (cft) at ({2+cos(45)},{sin(45)}){};
        \node[flavourb, label=right:$C_1$] (cfb) at ({2+cos(45)},{-sin(45)}){};

        \draw[-] (d1l)--(c1l)--(d1t)--(cft) (c1l)--(d1b)--(cfb);
    \end{tikzpicture}\label{eq:JunctionExample1}
\end{equation}The refined Coulomb branch Hilbert series may be evaluated by gluing together Hall-Littlewood polynomials. The refined Hilbert series is too cumbersome to present so the unrefined Hilbert series is presented for brevity \begin{align}
    \hs\left[\mathcal C\left(\text{\Quiver{eq:JunctionExample1}}\right)\right]&=\frac{1 + 4 t^2 + 22 t^4 + 46 t^6 + 77 t^8 + 92 t^{10} + 77 t^{12} + 46 t^{14} + 
 22 t^{16} + 4 t^{18} + t^{20}}{(1 - t^2)^4 (1 - t^4)^2  (1 - t^6)^2}\\\PL\left[\hs\left[\mathcal C\left(\text{\Quiver{eq:JunctionExample1}}\right)\right]\right]&=8 t^2 + 14 t^4 - 20 t^6 - 68 t^8 + O(t^{10})
\end{align} This Coulomb branch has no particular name as a symplectic singularity, although the global symmetry is identified as $\sorm(4)\times\urm(1)\times\urm(1)$.

The $\sorm(3)$ \hyperref[sec:rules]{quotient quiver subtraction} is shown in \Figref{fig:JunctionExample1} where each alignment of the $\sorm(3)$ quotient quiver gives $\sorm(2)$ with two flavours. Therefore the result of the hyper-Kähler quotient on the Coulomb branch of \Quiver{eq:JunctionExample1} is the union of the Coulomb branches of the two copies of $\sorm(2)$ with two flavours. The Coulomb branch of $\sorm(2)$ with two flavours is the Klein $A_3$ singularity, the intersection of the two identical quivers is trivial. The unrefined Hilbert series of the union of Coulomb branches is computed as \begin{equation}
    \hs\left[\mathcal C\left(\text{\Quiver{eq:JunctionExample1}}\right)///\sorm(3)\right]=\hs\left[A_3\cup A_3\right]=2\times \frac{1-t^8}{(1-t^2)(1-t^4)^2}-1=\frac{1+t^2+3t^4-t^6}{(1-t^2)(1-t^4)}
\end{equation}
\begin{figure}[h!]
    \centering
    \begin{subfigure}{0.45\textwidth}
        \begin{tikzpicture}
        \node[gauger, label=below:$D_1$] (d1l) at (0,0){};
        \node[gaugeb, label=below:$C_1$] (c1l) at (1,0){};
        \node[gauger, label=above:$D_1$] (d1t) at ({1+cos(45)},{sin(45)}){};
        \node[gauger, label=above:$D_1$] (d1b) at ({1+cos(45)},{-sin(45)}){};
        \node[flavourb, label=right:$C_1$] (cft) at ({2+cos(45)},{sin(45)}){};
        \node[flavourb, label=right:$C_1$] (cfb) at ({2+cos(45)},{-sin(45)}){};

        \draw[-] (d1l)--(c1l)--(d1t)--(cft) (c1l)--(d1b)--(cfb);

        \node[gauger, label=right:$D_1$] (D1rs) at ({1+cos(45)},{-2+sin(45)}){};
        \node[gaugeb, label=below:$C_1$] (C1s) at (1,-2){};
        \node[gauger, label=below:$D_1$] (D1ls) at (0,-2){};

        \draw[-] (D1rs)--(C1s)--(D1ls);

        \node[] (minus) at (-1,-2){$-$};

        \node[flavourb,label=above:$C_2$] (C2f) at (2,-4){};
        \node[gauger, label=below:$D_1$] (D1) at (2,-5){};

        \draw[-] (C2f)--(D1);

        \end{tikzpicture}
        \caption{}
        \label{}
    \end{subfigure}
    \begin{subfigure}{0.45\textwidth}
        \begin{tikzpicture}
      \node[gauger, label=below:$D_1$] (d1l) at (0,0){};
        \node[gaugeb, label=below:$C_1$] (c1l) at (1,0){};
        \node[gauger, label=above:$D_1$] (d1t) at ({1+cos(45)},{sin(45)}){};
        \node[gauger, label=above:$D_1$] (d1b) at ({1+cos(45)},{-sin(45)}){};
        \node[flavourb, label=right:$C_1$] (cft) at ({2+cos(45)},{sin(45)}){};
        \node[flavourb, label=right:$C_1$] (cfb) at ({2+cos(45)},{-sin(45)}){};

        \draw[-] (d1l)--(c1l)--(d1t)--(cft) (c1l)--(d1b)--(cfb);

        \node[gauger, label=right:$D_1$] (D1rs) at ({1+cos(45)},{-2-sin(45)}){};
        \node[gaugeb, label=below:$C_1$] (C1s) at (1,-2){};
        \node[gauger, label=below:$D_1$] (D1ls) at (0,-2){};

        \draw[-] (D1rs)--(C1s)--(D1ls);

        \node[] (minus) at (-1,-2){$-$};

        \node[flavourb,label=above:$C_2$] (C2f) at (2,-4){};
        \node[gauger, label=below:$D_1$] (D1) at (2,-5){};

        \draw[-] (C2f)--(D1);
        \end{tikzpicture}
        \caption{}
        \label{}
    \end{subfigure}
    \caption{$\sorm(3)$ \hyperref[sec:rules]{quotient quiver subtraction} on \Quiver{eq:JunctionExample1} to produce a union of two copies of $\sorm(2)$ with 4 flavours.}
    \label{fig:JunctionExample1}
\end{figure}
\paragraph{Example 2}
A further test of the junction rule would be when different quivers arise from different alignments. Consider the following quiver 
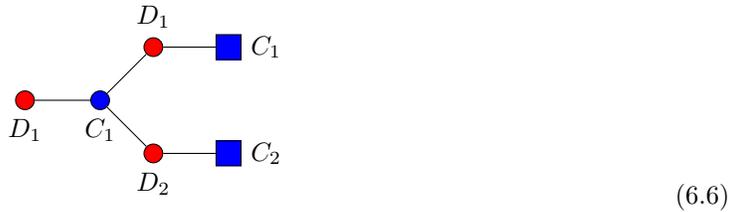
\begin{equation}
     \begin{tikzpicture}
        \node[gauger, label=below:$D_1$] (d1l) at (0,0){};
        \node[gaugeb, label=below:$C_1$] (c1l) at (1,0){};
        \node[gauger, label=above:$D_1$] (d1t) at ({1+cos(45)},{sin(45)}){};
        \node[gauger, label=below:$D_2$] (d1b) at ({1+cos(45)},{-sin(45)}){};
        \node[flavourb, label=right:$C_1$] (cft) at ({2+cos(45)},{sin(45)}){};
        \node[flavourb, label=right:$C_2$] (cfb) at ({2+cos(45)},{-sin(45)}){};

        \draw[-] (d1l)--(c1l)--(d1t)--(cft) (c1l)--(d1b)--(cfb);
    \end{tikzpicture}\label{eq:JunctionExample2}
\end{equation} whose Coulomb branch Hilbert series is evaluated as \begin{align}
    \hsC{eq:JunctionExample2}&=\frac{(1 + t^4) (1 + t^2 + 9 t^4 + 20 t^6 + 35 t^8 + 49 t^{10} + 60 t^{12} + 
   49 t^{14} + 35 t^{16} + 20 t^{18} + 9 t^{20} + t^{22} + t^{24})}{(1 - 
   t^2)^4 (1 - t^4)^3  (1 - t^6)^3}\\\PL\left[ \hsC{eq:JunctionExample2}\right]&=5 t^2 + 12 t^4 + 14 t^6 - 22 t^8 - 74 t^{10} - 7 t^{12} + O(t^{14})
\end{align}

The Coulomb branch global symmetry of \Quiver{eq:JunctionExample2} is $\sorm(3)\times\urm(1)\times\urm(1)$. Partial refinement of the Coulomb branch global symmetry is once again possible due to the gluing of Hall-Littlewood functions.

Subtraction of the $\sorm(3)$ quotient quiver is shown in \Figref{fig:JunctionExample2} with two possible alignments giving \Quiver{fig:JunctionExample2Q1}, the product quiver \Quiver{fig:JunctionExample2Q2}, and their intersection \Quiver{fig:JunctionExample2QInt} which is related by an Klein $A_3$ Kraft-Procesi transition. The Coulomb branch of \Quiver{fig:JunctionExample2Q1} is $\mathcal S^{\sorm(7)}_{\mathcal N,(3^2,1)}$. The Coulomb branch of $(B_1)-[C_2]$ is the Klein $A_3$ singularity and the Coulomb branch of $(D_1)-[C_2]$ is also the Klein $A_3$ singularity.

Although the hyper-Kähler quotient is checked at the level of the (partially) refined Hilbert series, the unrefined Hilbert series are presented for brevity to demonstrate agreement.\begin{equation}
    \hs\left[\mathcal C\left(\text{\Quiver{eq:JunctionExample2}}\right)///\sorm(3)\right]=\hs\left[\mathcal S^{\sorm(7)}_{\mathcal N,(3^2,1)}\cup \left(A_3\times A_3\right)\right]=\frac{(1 - t^8) (1 + 3 t^4 - t^6)}{(1 - t^2)^2 (1 - t^4)^3}
\end{equation}

\begin{figure}[h!]
    \centering
    \begin{subfigure}{0.45\textwidth}
    \centering
        \begin{tikzpicture}
        \node[gauger, label=below:$D_1$] (d1l) at (0,0){};
        \node[gaugeb, label=below:$C_1$] (c1l) at (1,0){};
        \node[gauger, label=above:$D_1$] (d1t) at ({1+cos(45)},{sin(45)}){};
        \node[gauger, label=above:$D_2$] (d1b) at ({1+cos(45)},{-sin(45)}){};
        \node[flavourb, label=right:$C_1$] (cft) at ({2+cos(45)},{sin(45)}){};
        \node[flavourb, label=right:$C_2$] (cfb) at ({2+cos(45)},{-sin(45)}){};

        \draw[-] (d1l)--(c1l)--(d1t)--(cft) (c1l)--(d1b)--(cfb);

        \node[gauger, label=right:$D_1$] (D1rs) at ({1+cos(45)},{-2+sin(45)}){};
        \node[gaugeb, label=below:$C_1$] (C1s) at (1,-2){};
        \node[gauger, label=below:$D_1$] (D1ls) at (0,-2){};

        \draw[-] (D1rs)--(C1s)--(D1ls);

        \node[] (minus) at (-1,-2){$-$};

        \node[flavourb,label=above:$C_3$] (C2f) at (2,-4){};
        \node[gauger, label=below:$D_2$] (D1) at (2,-5){};

        \draw[-] (C2f)--(D1);

        \end{tikzpicture}
        \caption{}
        \label{fig:JunctionExample2Q1}
    \end{subfigure}
    \begin{subfigure}{0.45\textwidth}
    \centering
        \begin{tikzpicture}
      \node[gauger, label=below:$D_1$] (d1l) at (0,0){};
        \node[gaugeb, label=below:$C_1$] (c1l) at (1,0){};
        \node[gauger, label=above:$D_1$] (d1t) at ({1+cos(45)},{sin(45)}){};
        \node[gauger, label=above:$D_2$] (d1b) at ({1+cos(45)},{-sin(45)}){};
        \node[flavourb, label=right:$C_1$] (cft) at ({2+cos(45)},{sin(45)}){};
        \node[flavourb, label=right:$C_2$] (cfb) at ({2+cos(45)},{-sin(45)}){};

        \draw[-] (d1l)--(c1l)--(d1t)--(cft) (c1l)--(d1b)--(cfb);

        \node[gauger, label=right:$D_1$] (D1rs) at ({1+cos(45)},{-2-sin(45)}){};
        \node[gaugeb, label=below:$C_1$] (C1s) at (1,-2){};
        \node[gauger, label=below:$D_1$] (D1ls) at (0,-2){};

        \draw[-] (D1rs)--(C1s)--(D1ls);

        \node[] (minus) at (-1,-2){$-$};

        \node[flavourb,label=above:$C_2$] (C2f) at (2,-4){};
        \node[gauger, label=below:$B_1$] (B1) at (2,-5){};

        \node[flavourb,label=above:$C_2$] (C2f2) at (3,-4){};
        \node[gauger, label=below:$D_1$] (D1) at (3,-5){};

        \draw[-] (C2f)--(B1);
        \draw[-] (C2f2)--(D1);
        \end{tikzpicture}
        \caption{}
        \label{fig:JunctionExample2Q2}
    \end{subfigure}
    \begin{subfigure}{\textwidth}
    \centering
    \begin{tikzpicture}
    \node[gauger, label=below:$B_1$] (b1) at (0,0){};
    \node[flavourb, label=above:$C_2$] (c2f) at (0,1){};
    \draw[-] (b1)--(c2f);
    \end{tikzpicture}
    \caption{}
    \label{fig:JunctionExample2QInt}
    
    \end{subfigure}
    \caption{$\sorm(3)$ \hyperref[sec:rules]{quotient quiver subtraction} on \Quiver{eq:JunctionExample2} to produce \Quiver{fig:JunctionExample2Q1}, the product quiver \Quiver{fig:JunctionExample2Q2}, and their intersection \Quiver{fig:JunctionExample2QInt}.}
    \label{fig:JunctionExample2}
\end{figure}
\paragraph{Example 3}
A further test of the junction rule would be when there are more than two possible alignments. Consider the following quiver \begin{equation}
     \begin{tikzpicture}
        \node[gauger, label=below:$D_1$] (d1l) at (0,0){};
        \node[gaugeb, label=below:$C_1$] (c1l) at (1,0){};
        \node[gauger, label=above:$D_1$] (d1t) at ({1+cos(45)},{sin(45)}){};
        \node[gauger, label=below:$D_1$] (d1b) at ({1+cos(45)},{-sin(45)}){};
        \node[flavourb, label=right:$C_1$] (cft) at ({2+cos(45)},{sin(45)}){};
        \node[flavourb, label=right:$C_1$] (cfb) at ({2+cos(45)},{-sin(45)}){};

        \node[gauger, label=below:$D_1$] (d1r) at (2,0){};
        \node[flavourb, label=right:$C_1$] (c1fr) at (3,0){};

        \draw[-] (d1l)--(c1l)--(d1t)--(cft) (c1l)--(d1b)--(cfb) (c1l)--(d1r)--(c1fr);
    \end{tikzpicture}\label{eq:JunctionExample3}
\end{equation} whose Coulomb branch Hilbert series is computed as \begin{align}
    \hsC{eq:JunctionExample3}=\frac{\left(\begin{aligned}1 &+ 3 t^2 + 14 t^4 + 49 t^6 + 150 t^8 + 339 t^{10} + 689 t^{12} + 
   1218 t^{14} \\&+ 1925 t^{16} + 2705 t^{18} + 3437 t^{20} + 3966 t^{22} + 4152 t^{24} + \cdots + 
   t^{48}\end{aligned}\right)}{(1 - t^2)^3 (1 - t^4) (1 - t^6)^2 (1 - t^8)^2  (1 - t^{10})^2}
\end{align}The Coulomb branch global symmetry of \Quiver{eq:JunctionExample3} is $\sorm(3)\times\urm(1)\times\urm(1)\times\urm(1)$. Partial refinement of the Coulomb branch global symmetry is once again possible due to the gluing of Hall-Littlewood functions.

Subtraction of just one alignment of the $\sorm(3)$ quotient quiver is shown in \Figref{fig:JunctionExample3} producing the product quiver \Quiver{fig:JunctionExample3}. The full result of the $\sorm(3)$ \hyperref[sec:rules]{quotient quiver subtraction} is a union of three copies of the product quiver \Quiver{fig:JunctionExample3}. The Coulomb branch of each $(D_1)-[C_2]$ is Klein $A_3$.

Although the hyper-Kähler quotient is checked at the level of the (partially) refined Hilbert series, the unrefined Hilbert series are presented for brevity to demonstrate agreement.\begin{equation}
    \hs\left[\mathcal C\left(\text{\Quiver{eq:JunctionExample3}}\right)///\sorm(3)\right]=\hs\left[\cup^3\left(A_3\times A_3\right)\right]=\frac{1 + t^2 + 5 t^4 + 4 t^6 + 5 t^8 - 
 5 t^{10} + t^{12}}{(1 - t^2)^2 (1 - t^4)^2}
\end{equation}
As the above is the Hilbert series of a union of three quivers, the two-way and three-way intersections appear with a negative and positive sign respectively in the union of cones formula.
\begin{figure}[h!]
    \centering
        \begin{tikzpicture}
        \node[gauger, label=below:$D_1$] (d1l) at (0,0){};
        \node[gaugeb, label=below:$C_1$] (c1l) at (1,0){};
        \node[gauger, label=above:$D_1$] (d1t) at ({1+cos(45)},{sin(45)}){};
        \node[gauger, label=below:$D_1$] (d1b) at ({1+cos(45)},{-sin(45)}){};
        \node[flavourb, label=right:$C_1$] (cft) at ({2+cos(45)},{sin(45)}){};
        \node[flavourb, label=right:$C_1$] (cfb) at ({2+cos(45)},{-sin(45)}){};

        \node[gauger, label=below:$D_1$] (d1r) at (2,0){};
        \node[flavourb, label=right:$C_1$] (c1fr) at (3,0){};

        \draw[-] (d1l)--(c1l)--(d1t)--(cft) (c1l)--(d1b)--(cfb) (c1l)--(d1r)--(c1fr);

        \node[gauger, label=right:$D_1$] (D1rs) at ({1+cos(45)},{-2.5+sin(45)}){};
        \node[gaugeb, label=below:$C_1$] (C1s) at (1,-2.5){};
        \node[gauger, label=below:$D_1$] (D1ls) at (0,-2.5){};

        \draw[-] (D1rs)--(C1s)--(D1ls);

        \node[] (minus) at (-1,-2.5){$-$};

        \node[flavourb,label=above:$C_2$] (C2f) at (2,-4){};
        \node[gauger, label=below:$D_1$] (D1) at (2,-5){};

        \draw[-] (C2f)--(D1);

        \node[flavourb,label=above:$C_2$] (C2f2) at (1,-4){};
        \node[gauger, label=below:$D_1$] (D12) at (1,-5){};

        \draw[-] (C2f2)--(D12);

        \end{tikzpicture}
    \caption{One alignment of the $\sorm(3)$ \hyperref[sec:rules]{quotient quiver subtraction} on \Quiver{eq:JunctionExample3} to produce the product quiver \Quiver{fig:JunctionExample3}. The result of the algorithm is a union of three copies of \Quiver{fig:JunctionExample3}.}
    \label{fig:JunctionExample3}
\end{figure}
\paragraph{Example 4}
The examples tested so far involve subtraction of the $\sorm(3)$ quotient quiver, here subtraction of the $\sorm(4)$ quotient quiver is studied in an example where the junction rule is encountered.

Consider the following quiver \begin{equation}
    \begin{tikzpicture}
        \node[gauger, label=below:$D_1$] (d1l) at (0,0){};
        \node[gaugeb, label=below:$C_1$] (c1l) at (1,0){};
        \node[gauger, label=below:$D_2$] (d2l) at (2,0){};
        \node[gaugeb, label=below:$C_2$] (c2l) at (3,0){};
        \node[gauger, label=above:$D_2$] (d2t) at ({3+cos(45)},{sin(45)}){};
        \node[gauger, label=below:$D_2$] (d2b) at ({3+cos(45)},{-sin(45)}){};
        \node[flavourb, label=right:$C_1$] (c1ft) at ({4+cos(45)},{sin(45)}){};
        \node[flavourb, label=right:$C_1$] (c1fb) at ({4+cos(45)},{-sin(45)}){};

        \draw[-] (d1l)--(c1l)--(d2l)--(c2l)--(d2t)--(c1ft) (c2l)--(d2b)--(c1fb);
    \end{tikzpicture}\label{eq:JunctionExample4}
\end{equation}The Coulomb branch may be evaluated with refinement by gluing together Hall-Littlewood polynomials. For brevity only the unrefined Hilbert series is presented as \begin{align}
    \hsC{eq:JunctionExample4}&=\frac{\left(\begin{aligned}1 &+ 4 t^2 + 30 t^4 + 128 t^6 + 546 t^8 + 1846 t^{10} + 5414 t^{12} + 
   13328 t^{14} + 28279 t^{16} \\&+ 51672 t^{18} + 82188 t^{20} + 114024 t^{22} + 
   138678 t^{24} + 147948 t^{26} + \cdots +
   t^{52}\end{aligned}\right)}{(1 - t^2)^8  (1 - t^4)^8 (1 - t^6)^4}\\\PL\left[ \hsC{eq:JunctionExample4}\right]&=12 t^2 + 28 t^4 + 32 t^6 - 11 t^8 - 246 t^{10} - 902 t^{12} + O(t^{14})
\end{align}the global symmetry is identified as $\sorm(5)\times\urm(1)\times\urm(1)$.

Subtraction of the $\sorm(4)$ quotient quiver on \Quiver{eq:JunctionExample4} is shown in \Figref{fig:JunctionExample4} to produce two identical quivers \Quiver{fig:JunctionExample4Q1} and \Quiver{fig:JunctionExample4Q2}. Their intersection follows from a Klein $A_3$ Kraft-Procesi transition resulting in \Quiver{fig:JunctionExample4QInt}. The Coulomb branches of these quivers have no particular name however their Hilbert series are presented below \begin{align}
    \hsC{fig:JunctionExample4Q1}&=\frac{1 + 5 t^4 + 7 t^6 + 8 t^8 + 14 t^{10} + 8 t^{12} + 7 t^{14} + 
 5 t^{16} + t^{20}}{(1 - t^2)^2 (1 - t^4)^6}\\
 \PL\left[\hsC{fig:JunctionExample4Q1}\right]&=2 t^2 + 11 t^4 + 7 t^6 - 7 t^8 - 21 t^{10} - 20 t^{12} + O(t^{14})\\
 \hsC{fig:JunctionExample4QInt}&=\frac{1 + 5 t^4 + 4 t^6 + 4 t^8 + 5 t^{10} + t^{14}}{(1 - t^2)^2 (1 - t^4)^4}\\\PL\left[\hsC{fig:JunctionExample4QInt}\right]&=2 t^2 + 9 t^4 + 4 t^6 - 11 t^8 - 15 t^{10} + 10t^{12}+O(t^{14})
\end{align}

These unrefined Hilbert series hence show that an $\sorm(4)$ hyper-Kähler quotient on the Coulomb branch of \Quiver{eq:JunctionExample4} agrees with that arising from the \hyperref[sec:rules]{quotient quiver subtraction} algorithm,\begin{equation}
    \hs\left[\mathcal C\left(\text{\Quiver{eq:JunctionExample4}}\right)///\sorm(4)\right]=\hs\left[\mathcal C\left(\text{\Quiver{fig:JunctionExample4Q1}}\right)\cup \mathcal C\left(\text{\Quiver{fig:JunctionExample4Q1}}\right)\right]=\frac{(1 + t^4) (1 - 2 t^2 + 9 t^4 - 6 t^6 + 18 t^8 - 9 t^{10} + 4 t^{12} - 
   t^{14})}{(1 - t^2)^4 (1 - t^4)^4}
\end{equation}

\begin{figure}[h!]
    \centering
    \begin{subfigure}{0.45\textwidth}
    \centering
        \begin{tikzpicture}
        \node[gauger, label=below:$D_1$] (d1l) at (0,0){};
        \node[gaugeb, label=below:$C_1$] (c1l) at (1,0){};
        \node[gauger, label=below:$D_2$] (d2l) at (2,0){};
        \node[gaugeb, label=below:$C_2$] (c2l) at (3,0){};
        \node[gauger, label=above:$D_2$] (d2t) at ({3+cos(45)},{sin(45)}){};
        \node[gauger, label=above:$D_2$] (d2b) at ({3+cos(45)},{-sin(45)}){};
        \node[flavourb, label=right:$C_1$] (c1ft) at ({4+cos(45)},{sin(45)}){};
        \node[flavourb, label=right:$C_1$] (c1fb) at ({4+cos(45)},{-sin(45)}){};

        \draw[-] (d1l)--(c1l)--(d2l)--(c2l)--(d2t)--(c1ft) (c2l)--(d2b)--(c1fb);

        \node[gauger, label=below:$D_1$] (d1ls) at (0,-2){};
        \node[gaugeb, label=below:$C_1$] (c1ls) at (1,-2){};
        \node[gauger, label=below:$D_2$] (d2ls) at (2,-2){};
        \node[gaugeb, label=below:$C_1$] (c1rs) at (3,-2){};
        \node[gauger, label=right:$D_1$] (d1rs) at ({3+cos(45)},{-2+sin(45)}){};

        \draw[-] (d1ls)--(c1ls)--(d2ls)--(c1rs)--(d1rs){};

        \node[] (minus) at (-1,-2){$-$};

        \node[gaugeb, label=below:$C_1$] (C1) at (3,-4){};
        \node[gauger, label=above:$B_1$] (B1) at ({3+cos(45)},{-4+sin(45)}){};
        \node[gauger, label=below:$D_2$] (D2) at ({3+cos(45)},{-4-sin(45)}){};
        \node[flavourb, label=right:$C_1$] (C1f) at ({4+cos(45)},{-4+sin(45)}){};
        \node[flavourb, label=right:$C_2$] (C2f) at ({4+cos(45)},{-4-sin(45)}){};
        \node[flavourr, label=left:$B_0$] (B0f) at (3,-3){};

        \draw[-] (B0f)--(C1)--(B1)--(C1f) (C1)--(D2)--(C2f);

        \end{tikzpicture}
        \caption{}
        \label{fig:JunctionExample4Q1}
    \end{subfigure}
    \begin{subfigure}{0.45\textwidth}
    \centering
        \begin{tikzpicture}
      \node[gauger, label=below:$D_1$] (d1l) at (0,0){};
        \node[gaugeb, label=below:$C_1$] (c1l) at (1,0){};
        \node[gauger, label=below:$D_2$] (d2l) at (2,0){};
        \node[gaugeb, label=below:$C_2$] (c2l) at (3,0){};
        \node[gauger, label=above:$D_2$] (d2t) at ({3+cos(45)},{sin(45)}){};
        \node[gauger, label=above:$D_2$] (d2b) at ({3+cos(45)},{-sin(45)}){};
        \node[flavourb, label=right:$C_1$] (c1ft) at ({4+cos(45)},{sin(45)}){};
        \node[flavourb, label=right:$C_1$] (c1fb) at ({4+cos(45)},{-sin(45)}){};

        \draw[-] (d1l)--(c1l)--(d2l)--(c2l)--(d2t)--(c1ft) (c2l)--(d2b)--(c1fb);

        \node[gauger, label=below:$D_1$] (d1ls) at (0,-2){};
        \node[gaugeb, label=below:$C_1$] (c1ls) at (1,-2){};
        \node[gauger, label=below:$D_2$] (d2ls) at (2,-2){};
        \node[gaugeb, label=below:$C_1$] (c1rs) at (3,-2){};
        \node[gauger, label=right:$D_1$] (d1rs) at ({3+cos(45)},{-2-sin(45)}){};

        \draw[-] (d1ls)--(c1ls)--(d2ls)--(c1rs)--(d1rs){};

        \node[] (minus) at (-1,-2){$-$};

        \node[gaugeb, label=below:$C_1$] (C1) at (3,-4){};
        \node[gauger, label=below:$B_1$] (B1) at ({3+cos(45)},{-4-sin(45)}){};
        \node[gauger, label=below:$D_2$] (D2) at ({3+cos(45)},{-4+sin(45)}){};
        \node[flavourb, label=right:$C_1$] (C1f) at ({4+cos(45)},{-4-sin(45)}){};
        \node[flavourb, label=right:$C_2$] (C2f) at ({4+cos(45)},{-4+sin(45)}){};
        \node[flavourr, label=left:$B_0$] (B0f) at (3,-3){};

        \draw[-] (B0f)--(C1)--(B1)--(C1f) (C1)--(D2)--(C2f);
        \end{tikzpicture}
        \caption{}
        \label{fig:JunctionExample4Q2}
    \end{subfigure}
    \begin{subfigure}{\textwidth}
    \centering
    \begin{tikzpicture}
    \node[gauger, label=below:$B_1$] (b1l) at (0,0){};
    \node[gaugeb, label=below:$C_1$] (c1) at (1,0){};
    \node[gauger, label=below:$B_1$] (b1r) at (2,0){};
    \node[flavourb, label=left:$C_1$] (C1fl) at (0,1){};
    \node[flavourr, label=above:$D_1$] (D1f) at (1,1){};
    \node[flavourb, label=right:$C_1$] (C1fr) at (2,1){};

    \draw[-] (C1fl)--(b1l)--(c1)--(b1r)--(C1fr) (D1f)--(c1);
    \end{tikzpicture}
    \caption{}
    \label{fig:JunctionExample4QInt}
    
    \end{subfigure}
    \caption{$\sorm(4)$ \hyperref[sec:rules]{quotient quiver subtraction} on \Quiver{eq:JunctionExample4} to produce \Quiver{fig:JunctionExample4Q1} with each alignment and their intersection \Quiver{fig:JunctionExample4QInt}.}
    \label{fig:JunctionExample4}
\end{figure}

The hyper-Kähler quotient is performed with the $\sorm(5)\hookleftarrow\sorm(4)$ embedding which decomposes the $\sorm(5)$ vector as\begin{equation}
    [1,0]_{\sorm(5)}\rightarrow [1,1]_{\sorm(4)}+[0,0]_{\sorm(4)}
\end{equation}

\section{Outlook}
\label{sec:outlook}
This paper extends the prescription of orthosymplectic \hyperref[sec:rules]{quotient quiver subtraction} to the case of framed orthosymplectic quivers, following work on the unitary \cite{Hanany:2023tvn} and unframed orthosymplectic \cite{Bennett:2024llh} cases. The quotient quivers for framed orthosymplectic quivers take the form of magnetic quivers for class $\mathcal S$ theories on a cylinder with maximal ($\mathbb Z_2$-twisted) D-type punctures or with maximal $\mathbb Z_2$-twisted A-type punctures. These gauge $\sorm(2n+1)$, $\sorm(2n)$, and $\sprm(n)$ subgroups of the Coulomb branch global symmetry, with complete Higgsing, respectively. There are now quotient quivers that originate from class $\mathcal S$ for all classical groups. Owing to computational challenges for Coulomb branch Hilbert series for quivers with negatively balanced gauge nodes, only the cases for $\sorm(2n+1)$ and $\sorm(2n)$ are studied in detail here. Unfortunately, attempts at performing $\sprm(k)$ quotient quiver subtraction does not yield quivers with computable moduli spaces. Obtaining evidence for the $\sprm(k)$ case is an open problem.

An important conceptual point regarding the combinatorics of orthosymplectic quivers is that the combinatorics for framed and unframed quivers is generically different. This manifests itself for quotient quiver subtraction as having different sets of quotient quivers for each case and also in some details of the two subtraction algorithms, addressed in Section \ref{subsec:comparison}. However, general lessons from this point can be applied to the future development of other combinatorial techniques on orthosymplectic quivers. One such example would be quiver subtraction algorithms to derive the stratification of the Coulomb branch or of the Higgs branch.





In contrast to the unframed case, the framed quotient quivers introduced in this work apply strictly for the classical groups. Whether there exist exceptional quotient quivers for framed theories is unknown and remains as an open question.

\acknowledgments
We thank Mohammad Akhond, Guillermo Arias-Tamargo, Michael Finkelberg, Rudolph Kalveks, Deshuo Liu, Lorenzo Mansi, and Hiraku Nakajima for useful discussions. The work of SB, AH, and GK is partially supported by STFC Consolidated Grants ST/T000791/1 and ST/X000575/1. The work of SB is supported by the STFC DTP research studentship grant ST/Y509231/1. The work of GK is supported by STFC DTP research studentship grant ST/X508433/1.
\bibliographystyle{JHEP}
\bibliography{references.bib}
\appendix
\section{An attempt at $\sprm(n)$ quotient quiver subtraction}
\label{sec:so(4n+2)_(4n+1)_flavours}

It is challenging to explicitly check $\sprm(n)$ \hyperref[sec:rules]{quotient quiver subtraction} since the quivers which admit such a subtraction necessarily have underbalanced gauge nodes. Despite this, an attempt of $\sprm(n)$ \hyperref[sec:rules]{quotient quiver subtraction} is presented -- at least schematically -- to demonstrate that it proceeds in a similar fashion to $\sorm(n)$ and $\surm(n)$ \hyperref[sec:rules]{quotient quiver subtraction}.

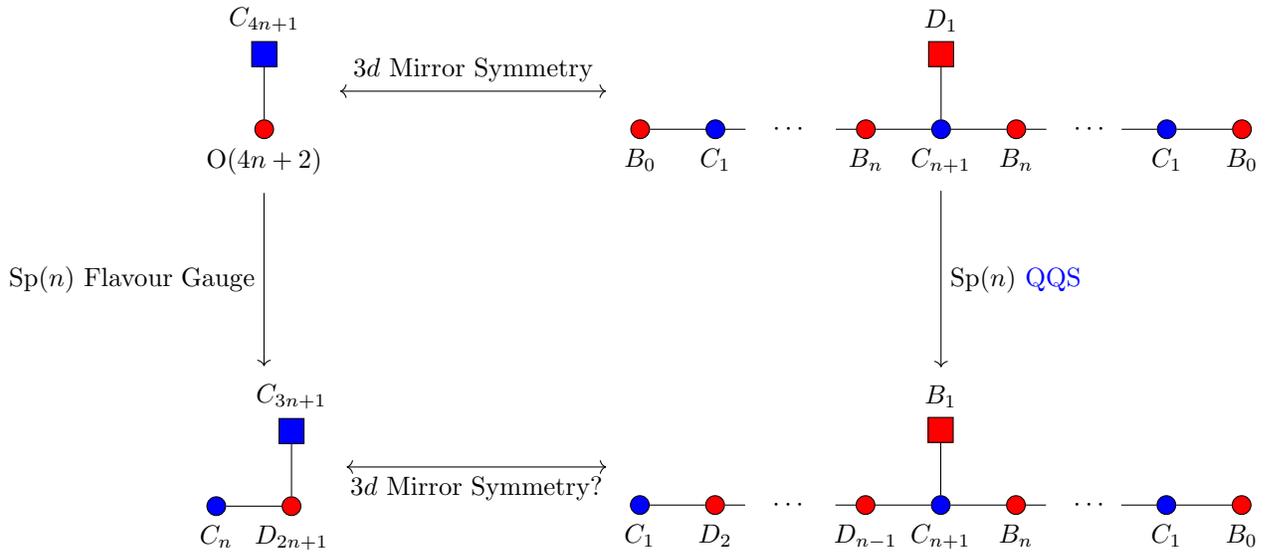
\begin{figure}[H]
\centering
\begin{tikzpicture}
    \node (a) at (0,0){$\begin{tikzpicture}
        \node (1) [gauger, label=below:$\orm(4n+2)$] at (0,0) {};
        \node (2) [flavourb, label=above:$C_{4n+1}$] at (0,1) {};
        \draw[-] (1)--(2);
    \end{tikzpicture}$};

    \node (b) at (9,0){$\begin{tikzpicture}
        \node (1) [gauger, label=below:{$B_0$}] at (0,0) {};
        \node (2) [gaugeb, label=below:{$C_1$}] at (1,0) {};
        \node (3) [gauger, label=below:{$B_{n}$}] at (3,0) {};
        \node (4) [gaugeb, label=below:{$C_{n+1}$}] at (4,0) {};
        \node (6) [flavourr, label=above:{$D_1$}] at (4,1) {};
        \node (7) [gauger, label=below:{$B_{n}$}] at (5,0) {};
        \node (8) [gaugeb, label=below:{$C_1$}] at (7,0) {};
        \node (9) [gauger, label=below:{$B_0$}] at (8,0) {};
        \draw (1) -- (2);
        \draw (3) -- (4) -- (6);
        \draw (2) -- (1.4,0);
        \draw (3) -- (2.6,0);
        \draw (4)--(7)--(5.4,0);
        \draw (6.4,0)--(8)--(9);
        \node at (2, 0) {$\cdots$} {};
        \node at (6, 0) {$\cdots$} {};
    \end{tikzpicture}$};

    \node (c) at (0,-5){$\begin{tikzpicture}
        \node (1) [gauger, label=below:$D_{2n+1}$] at (0,0) {};
        \node (2) [flavourb, label=above:$C_{3n+1}$] at (0,1) {};
        \node (3) [gaugeb, label=below:$C_n$] at (-1,0) {};
        \draw[-] (3)--(1)--(2);
    \end{tikzpicture}$};

    \node (d) at (9,-5){$\begin{tikzpicture}
        \node (1) [gaugeb, label=below:{$C_1$}] at (0,0) {};
        \node (2) [gauger, label=below:{$D_2$}] at (1,0) {};
        \node (3) [gauger, label=below:{$D_{n-1}$}] at (3,0) {};
        \node (4) [gaugeb, label=below:{$C_{n+1}$}] at (4,0) {};
        \node (6) [flavourr, label=above:{$B_1$}] at (4,1) {};
        \node (7) [gauger, label=below:{$B_{n}$}] at (5,0) {};
        \node (8) [gaugeb, label=below:{$C_1$}] at (7,0) {};
        \node (9) [gauger, label=below:{$B_0$}] at (8,0) {};
        \draw (1) -- (2);
        \draw (3) -- (4) -- (6);
        \draw (2) -- (1.4,0);
        \draw (3) -- (2.6,0);
        \draw (4)--(7)--(5.4,0);
        \draw (6.4,0)--(8)--(9);
        \node at (2, 0) {$\cdots$} {};
        \node at (6, 0) {$\cdots$} {};
    \end{tikzpicture}$};
    \draw[<->] (a)--(b)node[pos=0.5, above]{$3d$ Mirror Symmetry};
    \draw[<->] (c)--(d)node[pos=0.5, below]{$3d$ Mirror Symmetry?};
    \draw[->] (b)--(d)node[pos=0.5, right]{$\sprm(n)$ \hyperref[sec:rules]{QQS}};
    \draw[->] (a)--(c)node[pos=0.5, left]{$\sprm(n)$ Flavour Gauge};
    \end{tikzpicture}
    \caption{Commutative diagram showing the gauging of an $\sprm(n)$ flavour symmetry subgroup of $\orm(4n+1)$ SQCD with $4n+2$ flavours down the left column and $\sprm(n)$ \hyperref[sec:rules]{quotient quiver subtraction} (QQS) on the $3d$ mirror theory in the right column.}
    \label{quiv:cn_theory}
\end{figure}

Consider $\orm(4n+2)$ gauge theory with $4n+1$ flavours at finite coupling, given in the top left of \Figref{quiv:cn_theory}. Note that the choice of gauge group $\orm(4n+2)$ as opposed to $\sorm(4n+2)$ is essential for the $B_0$ nodes in the mirror theory to be gauged. Gauging an $\sprm(n)$ subgroup of the flavour symmetry results in the balanced $\sprm(n)\times\sorm(4n+2)$ gauge theory at the bottom left of \Figref{quiv:cn_theory} (note that the $D$-type gauge node is now $\sorm(4n+2)$ instead of $\orm(4n+2)$). The corresponding gauging on the mirror theory is given on the right-hand side of \Figref{quiv:cn_theory} through $\sprm(n)$ \hyperref[sec:rules]{quotient quiver subtraction}, where the flavour symmetry changes from $\sorm(2)$ to $\sorm(3)$. It remains to be seen whether the quivers in the bottom left and bottom right of \Figref{quiv:cn_theory} are $3d$ mirror.

\end{document}